\documentclass[aps,prfluid,twocolumn,superscriptaddress]{revtex4-2}
\usepackage{amsmath}
\usepackage{amssymb}
\usepackage{color}
\usepackage{graphicx}
\usepackage{epstopdf}
\usepackage{natbib}
\graphicspath{{figs/}}

\DeclareMathOperator{\arsinh}{arsinh}
\DeclareMathOperator{\artanh}{artanh}

\begin{document}
\title{Tuning diffusioosmosis of electrolyte solutions 
by hydrostatic pressure. }

\author{Elena F. Silkina}
\affiliation{Frumkin Institute of Physical Chemistry and Electrochemistry, Russian Academy of Sciences, 31-4 Leninsky Prospect, 119071 Moscow, Russia}

\author{Evgeny S. Asmolov}
\affiliation{Frumkin Institute of Physical Chemistry and Electrochemistry, Russian Academy of Sciences, 31-4 Leninsky Prospect, 119071 Moscow, Russia}

\author{Olga I. Vinogradova}
\email[Corresponding author: ]{oivinograd@yahoo.com}
\affiliation{Frumkin Institute of Physical Chemistry and Electrochemistry, Russian Academy of Sciences, 31-4 Leninsky Prospect, 119071 Moscow, Russia}
\begin{abstract}
When two reservoirs of a distinct salinity are connected by channels or pores, a fluid flow termed diffusio-osmotic is generated. This article investigates the flow emerging in an uniformly  charged long slit whose thickness exceeds the local Debye screening length. Attention is focussed on the role of hydrostatic pressure drop $\Delta p$  in establishing diffusioosmosis at a finite concentration difference. For a thick slit  we recover the  known formula  for a local  diffusio-osmotic slip over a single wall, which is   determined by the surface potential, salt concentration and its gradient. An equation for the global fluid flow rate $\mathcal{Q}$ is presented as a sum of the diffusio-osmotic and pressure-driven contributions. Although the diffusio-osmotic term itself remains unaffected by $\Delta p$, the nonlinear concentration and surface potential profiles along the slit, and consequently, the local slip velocity are dramatically modified.  We present an equation relating the local concentration to $\mathcal{Q}$  and employ it to derive an expression describing the surface potential variation in the slit. Since $\mathcal{Q}$ can easily be tuned by $\Delta p$, the variety of possible concentration and surface potential profiles becomes very rich.  Our theory provides a simple explanation of recent flow rate measurements and  shows that experimental data provide rather direct information about concentration and surface potential profiles in the uniformly charged slit. The relevance of our results for  sensing the salt dependence of surface potentials is discussed briefly.

\end{abstract}

\maketitle

\section{Introduction}\label{sec:intro}

A promising strategy to generate a fluid flow near a solid, inside micro- and nanofluidic channels or in the pores of permeable membranes is to employ the salinity gradients. Diffusioosmosis, which is the name given to this phenomenon whereby an electrolyte solution flows to high or low salinity~\cite{anderson1989colloid}, takes its origin in the electrostatic diffuse layers (EDLs) near a solid, where the tangential concentration gradient  and electric field
generate a force that sets the fluid in motion. In essence, such a stationary flow normally arises in response to the gradients of  both concentration and electric potential, but in contrast to a more familiar classical electroosmosis the field is not applied externally, but emerges spontaneously to provide zero current in an open circuit~\cite{deryagin1961}. The diffusioosmosis  provides an efficient way to induce and drive flows in micro- and nanofluidic devices, which does not require
an external energy supply. The only source of energy is the concentration gradient of solute, and we remark that it can also be easily induced by external stimuli, for example, by an appropriate illumination in the case of photosensitive ions~\cite{feldmann.d:2016}. In addition, being at the physical
origin of a diffusio-phoretic migration of colloid particles under salinity gradients, it  has long been invoked in the interpretation of their velocity~\cite{deryagin1961,ebel.jp:1988}.  During past years this includes a self-propulsion of ionic catalytic microswimmers, which opened a new
field of investigation with both fundamental and practical perspectives~\cite{de2020self,asmolov2022self,asmolov2022COCIS}.

In the development of colloid science,  experiments on diffusiophoresis  have played a great role indeed. This may be explained by the fact that
many relatively simple techniques for the tracking of particles and measuring their mobility are available. Another factor in favor of
diffusiophoresis certainly is that it represents a simple means to manipulate the charged colloids, by inducing their spreading and focusing~\cite{
abecassis.b:2009,wanunu.m:2010,shi.n:2016,rasmussen.mk:2020}. Since the diffusiophoresis is the reverse phenomenon from diffusioosmosis, these experiments  provided
indirect evidences of existence of the latter. The same concerns the experiments on passive and active colloids on the solid wall that can be
easily manipulated by the flow arising near it under the salinity gradient~\cite{
feldmann.d:2020,arya.p:2021,hardikar.av:2024,chakra.a:2025}. However, the diffusioosmosis itself that originates the viscous
stresses causing the particle diffusio-phoretic propulsion, being also significant for nanodevices,
remains much less investigated  experimentally. Direct experimental evidence of the generation of a diffusio-osmotic flow appeared only about a decade ago~\cite{lee.c:2014}. Namely, the fluid flow rate through a nanochannel (toward low salinity reservoirs) has been measured for several inorganic salt solutions.
Nevertheless,  information about such quantities, as ionic flux, local ion concentration and potential still cannot be obtained from direct experiment, so it is not surprising that the theoreticians have had free rein.

Quantitative understanding the diffusioosmosis of electrolyte solutions inside pores of a finite length in which the confining dimension (thickness or radius) is of the order of hundreds of nanometers or so, constitutes a fundamental and  challenging problem. The body of theoretical  work investigating this is much less than that for, say, electroosmosis due to its immense complexity. It has been indeed very difficult to give its attractive and simple interpretation. There is, however, a growing literature describing attempts to provide a satisfactory theory of diffusio-osmotic flow of salt solutions.

The equation by \citet{prieve1984motion} relates  the velocity of the diffusio-osmotic plug flow far from a single wall (termed the apparent slip velocity) to the (taken constant) surface potential and the derivative of the logarithm of local salt concentration, which is, in essence, unknown.
Despite an appealing simplicity of a model for electrolyte solutions flowing in  a slit, there have been relatively few theoretical studies of this configuration. In fact, generalising the single wall analysis~\cite{prieve1984motion} to the slit geometry is not so straightforward. The point is that beside a spontaneously emerging non-uniform electric field, to provide a zero current condition in a confined geometry the non-uniform pressure gradient is also induced inside the slit~\cite{peters.pb:2016}, which immediately complicates the analysis. In addition, the concentration drop between the ends is prescribed, but the local values of surface potential and concentration inside a slit are  established self-consistently and  have to be found, which is not an easy task. This is probably why the majority of analytical work simply assumed that the concentration distribution along the slit is linear~\cite{ma.hc:2006,keh.hj:2016,jing2018}. However, such an  approximation has limited applicability being justified only when the concentration drop is very small.

In an attempt to obtain a proper theoretical understanding of diffusio-osmotic flow at a finite concentration drop we have carried out a series of calculations for simple salt solutions confined in a slit connecting two bulk reservoirs of different salinity~\cite{asmolov.es:2025}. The calculations assumed that the walls of the slit obey hydrodynamic no-slip and electrostatic constant charge density boundary conditions, and that the slit itself is thick, which implies that EDLs are thin compared to its thickness. Our results showed that the assumption of a constant concentration gradient becomes unrealistic at a finite fluid flow rate. In such a case another factor, ``deformation'' of the linear concentration profiles by the fluid flow, comes into play~\cite{asmolov.es:2025}. Since the concentration profiles are strongly coupled to the emerging electric potentials and apparent slip velocity, they become a very important consideration in diffusioosmosis.
In the course of that work we gained the impression that several aspects of the diffusioosmosis of confined electrolyte solutions have been given insufficient attention. Here we focus on the impact of hydrostatic pressure difference between the reservoirs on diffusioosmosis in an uniformly charged thick slit of length $L$, whose walls are separated by distance $H \ll L$. The flow in such a long thick slit emerging due to concentration and hydrostatic pressure drops between the reservoirs was described by \citet{ault.jt:2019}, who assumed a constant surface potential, i.e. imposed an electrostatic boundary condition that is normally attributed to conductors. These authors, however, failed to derive an analytical solution for the case of insulating walls, where the surface potential varies along the slit. They also made no attempt to connect the fluid flow rate with the flux of ions.
In the present article we extend and generalise prior analysis~\cite{asmolov.es:2025,ault.jt:2019} by focusing on the role of a hydrostatic pressure drop in establishing the diffusio-osmotic flow   in uniformly charged slits.

Our paper is arranged as follows: We define our system and formulate the governing equations in Sec.~\ref{sec:model}. In Sec.~\ref{sec:potential} emphasis is placed on the surface potential $\phi_s$ and its concentration dependence. Section~\ref{sec:flowrate} describes the procedure for calculating the fluid flow rate $\mathcal{Q}$. We show that the presence of $\Delta p \neq 0$ results in a supplementary term in the expression for $\mathcal{Q}$, but does not affect a diffusio-osmotic contribution. Here we also clarify the origin for a salt specificity of the flow rate.  In Sec.~\ref{sec:fluxes2} the form of the concentration profiles is described; these depend on $\mathcal{Q}$, and, consequently, can be tuned by $\Delta p$. We also consider the flux of ions through the slit and relate it to the flow rate. Section~\ref{sec:midpotential} contains the results of calculation of surface potential profiles in the slit for various $\Delta p$. Results for diffusio-osmotic slip velocities are presented in Sec.~\ref{sec:slip_velocity}. We conclude in Sec.~\ref{sec:concl} with a discussion of our results and their possible relevance for further experiments and some potential applications. Appendix~\ref{a:1} makes contact with the experiments of~\citet{lee.c:2014}. Our theory provides a direct physical explanation of their results and infers the variation of concentration and surface potential inside the slit from these data.

\section{Model and governing equations}\label{sec:model}

\begin{figure}[h]
\includegraphics[width=0.6\columnwidth]{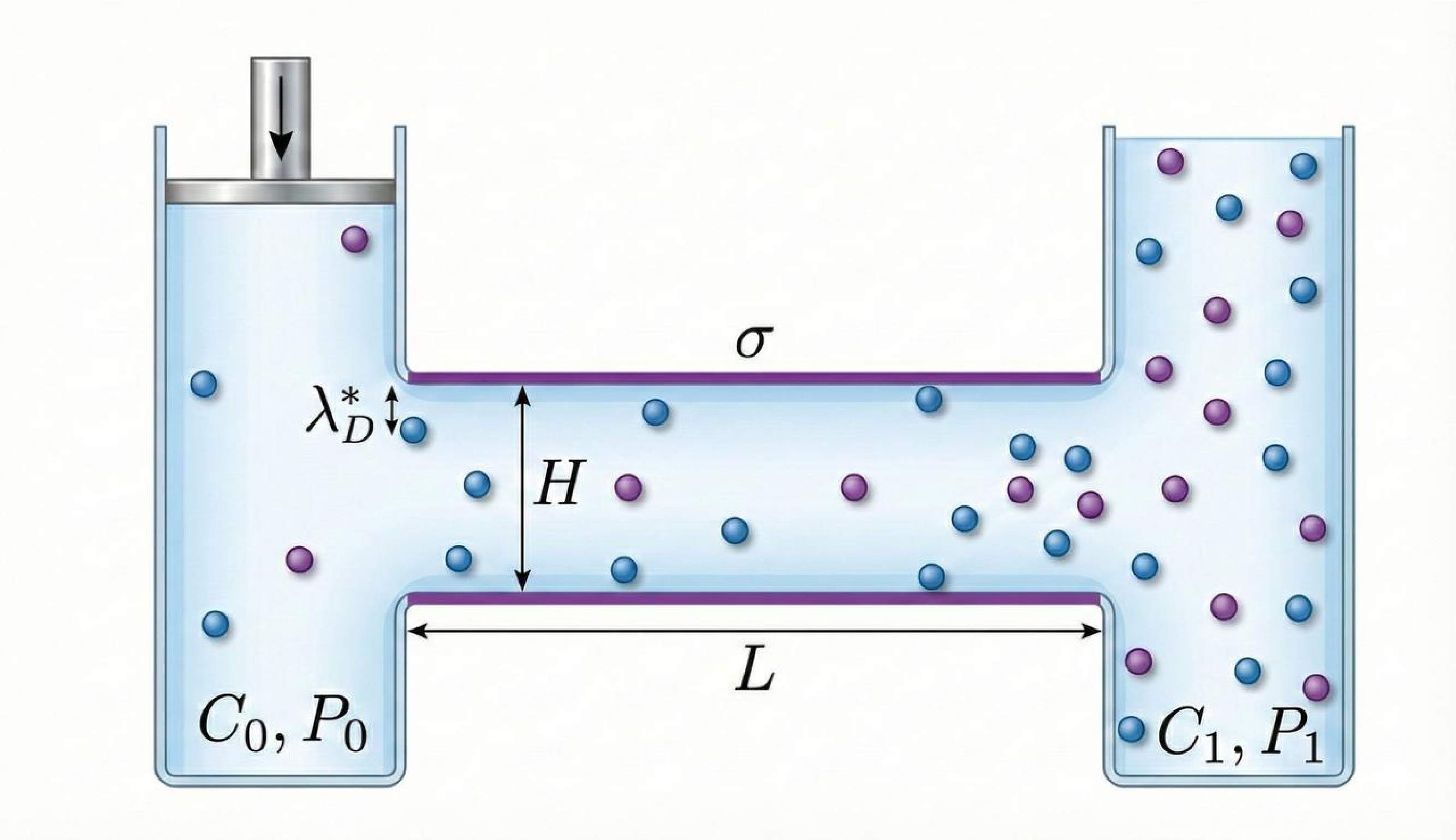}
\caption{Sketch of the microchannel of thickness $H$, length $L
\gg H$ and constant surface charge density $\sigma$ that connects the ``fresh'' (left) and ``salty'' (right) bulk electrolyte reservoirs of
concentrations $C_{0}$  and $C_{1}$ with hydrostatic pressures $P_{0}$  and $P_{1}$. The extension of electrostatic
diffuse layers, which is of the order of the Debye length $\protect\lambda_D \ll H$. It takes the upper value of $\lambda_D = \lambda_D^{\star}$ at $X=0$ and is reducing along the channel.   }
\label{fig:sketch}
\end{figure}

We consider a system sketched in Fig.~\ref{fig:sketch}. A planar slit of thickness $H$ and length $L$ is in contact with low (left) and  high (right) salinity  reservoirs of symmetric 1:1 salt solutions of concentrations (number densities of salt molecules) $C_0$ and $C_1$ at standard temperature $T$. Both solutions are of the same dynamic viscosity $\eta $ and permittivity $\varepsilon$, but their hydrostatic pressures $P_0$ and $P_1$ are generally unequal. The pressure drop $\Delta P = P_1 -P_0$ could be of any sign or vanish. There also exists an osmotic pressure drop between the reservoirs, which can always be evaluated using the classical
van't Hoff formula $\Delta \Pi = 2 k_B T (C_1 -C_0)$, where $k_{B}$ is the Boltzmann constant. Since $\Delta C = C_1 -C_0$ is positive, $\Delta \Pi$ can only be positive. Note that it normally is more convenient to use the concentration $\mathcal{C}[\rm{mol/l}]$, which is related to $C$ as $C \simeq N_A \times 10^3 \times \mathcal{C}$, where $N_A$ is Avogadro's number.  To get some idea of the orders of magnitude, the typical salinity contrast between sea and fresh water, $\Delta \mathcal{C} \simeq 1$ mol/l, yields $\Delta \Pi \simeq 5$ MPa (50 bar) at a standard temperature.

The walls of the slit represent nonconducting surfaces of a charge density $\sigma$, but rather than using $\sigma$ explicitly we here describe the surfaces by the Gouy-Chapman length, which is inversely proportional to the surface charge density (and may be positive or negative depending on its sign):
\begin{equation}
\ell _{GC}=\dfrac{e}{2\pi \sigma \ell _{B}},  \label{eq:LGC}
\end{equation}%
where $e$ is the elementary
positive charge and  $\ell _{B}=\dfrac{%
e^{2}}{\varepsilon k_{B}T}$ is the Bjerrum length characterising the solvent. The Bjerrum
length of water at $T \simeq 298$~K is equal to about $0.7$ nm leading to a useful relation $\ell _{GC} \mathrm{[nm]} \simeq \dfrac{36 \mathrm{[nm]}}{\sigma \mathrm{[mC/m^2]}}$.
The values of $\sigma$ of dielectric surfaces  inferred from electrokinetic experiments  and surface force  measurements have immense variability depending on their material, its preparation and modification, environment, and pH~\cite{behrens.sh:2001,butt.hj:2023,hartkamp.r:2018}. For example, some studies inferred a  surface charge about -4 mC/m$^2$ for silica surfaces that corresponds to $\ell_{GC} \simeq -10$ nm~\cite{rabinovich.yi:1982,ducker.wa:1991}, but what is essential for our work is that $\sigma$ can also be very high.
	\citet{yaroshchuk.a:2009} reported  $\sigma \simeq 16-25 $ mC/m$^2$ for nanoporous track-etched membranes, which yields $\ell_{GC} \simeq 2.3 -1.4$ nm.  \citet{stein.d:2004} inferred  $\sigma \simeq -60 $ mC/m$^2$ for silica channels that yields $\ell_{GC} \simeq -0.6$ nm. This is consistent with the upper (in magnitude) values of surface charge density of  silica obtained from other experiments~\cite{kitamura.a:1999,cerovic.ls:2002}. The surface charge density of mica can reach 40 mC/m$^2$, which is equivalent to $\ell_{GC} \simeq 0.9$ nm~\cite{claesson.pm:1986}.

The nonselective slit is fully permeable for both ionic species. In addition,  we assume that it is thick, which implies that a compensating
charge of the opposite sign and equal magnitude staying in the neighborhood of the charged walls is confined
in thin electrostatic diffuse layers (EDLs) near the walls, while an extended central region of the channel is
then, to the leading order, electro-neutral (``bulk'') in any cross-section. The local EDL thickness is of the order of the Debye screening length at a given cross-section and can be defined as $\lambda _{D} = \left[8\pi \ell _{B}C_m\right] ^{-1/2}$, where $C_m$ is the  concentration in the electro-neutral region (equal to that at the midplane). Clearly, on approaching the ``salty'' reservoir $\lambda _{D}$ reduces,  since the ``bulk'' concentration augments. The upper value of $\lambda_D$ is thus attained at the ``fresh'' end, where $C_m = C_0$, that yields
\begin{equation}\label{eq:DL2}
 \lambda^{\star}_{D} = \left[8\pi \ell _{B}C_0\right] ^{-1/2}.
\end{equation}
Thus, to fulfil the thick channel condition, it is enough to require $\lambda_{D}^{\star} \ll H$. Note that a useful relation to calculate the upper value of the Debye length is $\lambda^{\star}_D [\rm{nm}] \simeq \dfrac{0.305 [\rm{nm}]}{\sqrt{\mathcal{C}_0 [\rm{mol/l}]}}$.
Thus, on increasing $\mathcal{C}_0$ from $10^{-6}$ to $1$ mol/l the screening length reduces from about 300 down to 0.3 nm. For all specimen examples below we set $\mathcal{C}_0 = 10^{-3}$ mol/l, but vary $\mathcal{C}_1$ (and, therefore, the concentration drop $\Delta \mathcal{C}$). If so, the thick channel limit is expected when $H \geq 100$ nm.

For a given thickness $H$ and length $L$ the  solution inside the slit will adopt the configuration (i.e. distributions of the number density of the ionic species $C^{\pm}$, electrostatic potential $\Psi$ and hydrostatic pressure $P$) that minimizes the energy dissipation, i.e. corresponds to steady-state diffusioosmotic flow with the velocity $\mathbf{U}$. For a thick slit it is appropriate to consider the flow in the large electro-neutral region
by using apparent slip boundary conditions that can be applied at the imaginary smooth surface at a distance from the walls that is of the order of the EDL thickness.
Such an apparent condition substitutes the actual hydrodynamic one (here, no-slip) along the real wall and fully characterizes the fluid flow throughout the slit. Such a concept of ``apparent slip'' has been well justified for electrokinetic flows, including diffusioosmosis~\cite{prieve1984motion,asmolov.es:2025}, since it allows one to solve complex problems without tedious calculations. Our ultimate aim here is to obtain fundamental understanding of how a hydrostatic pressure drop $\Delta P$ between the ends of a thick slit would affect the apparent diffusio-osmotic slip inside it.

We place the origin of coordinates in the midplane of the slit (which is treated as unbounded in the $Y$ direction) at a junction with the low-salinity bath. The $Z$-axis is aligned across the slit, thus the walls  are located at $Z=\pm H/2$. The $X$-axis is defined along the channel, and $0\leq X\leq L$.

The stationary state of resulting diffusio-osmotic flow is described by the system of several
governing equations. The Nernst-Planck (or convection-diffusion) equation describes the conservation  of ionic
species at each point $(X,Z)$ inside the channel
\begin{equation}
\mathbf{\nabla }\cdot \mathbf{J}^{\pm }=0,  \label{NPH}
\end{equation}
where the ionic fluxes $\mathbf{J}^{\pm }$ of cations and anions are given by
\begin{equation}
\mathbf{J}^{\pm }=C^{\pm }\mathbf{U}+D^{\pm }\left( -\mathbf{\nabla }C^{\pm
}\mp \dfrac{e}{k_{B}T}C^{\pm }\mathbf{\nabla }\Psi \right).
\label{jd}
\end{equation}%
The first term in Eq.~\eqref{jd} is associated with the convective flux of ions induced by the flow, the second refers to  the diffusive drift relative to a solvent, and the third one is due to migration of ions in the emerging electric field.

The relation between the potential $\Psi$ and the charge density $\rho$  is
given by the Poisson equation:
\begin{equation}
\Delta \Psi =-\frac{\rho }{4\pi \varepsilon }=-\frac{e\left(
C^{+}-C^{-}\right) }{4\pi \varepsilon }.  \label{PEq}
\end{equation}%

The fluid flow satisfies the Stokes equations,
\begin{eqnarray}
\mathbf{\nabla \cdot U} &=&0\mathbf{,\quad }  \label{cont} \\
\eta \Delta \mathbf{U}-\mathbf{\nabla }P &=&\rho \mathbf{\nabla }\Psi .
\label{mom}
\end{eqnarray}%

It is clear that the theory for the diffusioosmosis requires, as input, the hydrodynamic and electrostatic boundary conditions at the channel walls. Here we impose classical no-slip hydrodynamic boundary condition ($U = 0$ at $Z = \pm H/2$) and consider non-conducting surfaces of charge density $\sigma$, which is constant. The latter implies that the surface charge density is independent on salt concentration, and, consequently, on $X$.

In the thick channel limit the local electrostatic potential in the slit is given by
\begin{equation}\label{eq:PSI}
 \Psi (X,Z) = \Psi_m (X) + \Phi (X,Z).
\end{equation}
Here the ``bulk'' term ($\Psi_m$) is supplemented by a ``surface'' term ($\Phi$) that represents the perturbation due to EDLs. Equation~\eqref{eq:PSI} implies that the emerging electric field that produces an electroosmotic flow is additively superimposed upon the field of the EDL.

To construct the solution of the system of Eqs.~\eqref{NPH}-\eqref{mom} it is convenient to define the dimensionless coordinates
\begin{equation*}
 x = \frac{X}{L},\quad z=\frac{2Z}{H}
\end{equation*}
that vary from $0$ to $1$. The dimensionless potentials are defined as
\begin{equation*}
\psi = \dfrac{e\Psi }{k_{B}T},\quad \phi = \dfrac{e\Phi }{k_{B}T}.
\end{equation*}
We also introduce the dimensionless variables~\cite%
{saville1977}
\begin{eqnarray*}
\mathbf{j}^{\pm } &=&\mathbf{J}^{\pm }\frac{2L}{C_{0}\left(
D^{+}+D^{-}\right) },\quad c^{\pm }=\frac{C^{\pm }}{C_{0}}, \\
\mathbf{u} &\mathbf{=}&\mathbf{U}\dfrac{4\pi \eta \ell_B L}{k_{B} T},\quad p=P\dfrac{\pi \ell_B H^{2}}{k_{B} T}
\end{eqnarray*}%

The ion fluxes given by \eqref{jd} can be rewritten in the dimensionless form as,
\begin{equation}
\mathbf{j}^{\pm }=\mathrm{Pe}c^{\pm }\mathbf{u}+\text{$\left( 1\pm \beta
\right) $}\left( -\mathbf{\nabla }c^{\pm }\mp c^{\pm }\mathbf{\nabla }\psi
\right) .  \label{NP1}
\end{equation}%
Here
\begin{equation}
\mathrm{Pe}=\dfrac{k_{B} T}{2\pi \eta \ell_B \left(
D^{+}+D^{-}\right) }  \label{Pe}
\end{equation}%
is the P\'{e}clet number that characterizes the ratio of the rate of
convection by that of diffusion. Note that $\mathrm{Pe}$ could also be interpreted as an inverse effective diffusion coefficient of a salt and that this (salt specific) parameter is always positive. The (also salt specific) factor $\beta $ is
defined in terms of the difference in diffusion constants of cations and
anions
\begin{equation}
\beta =\frac{D^{+}-D^{-}}{D^{+}+D^{-}},  \label{beta}
\end{equation}%
In such a definition $\beta $ is positive, if cations diffuse faster than
anions, and vice versa.

\begin{table}[h]
	\caption{Some typical values of $\beta$ and $\mathrm{Pe}$ calculated using the data on diffusion coefficients by~\citet{vanysek.r:2000}. }
	\label{table:beta}
	\begin{tabular}{|c|c|c|c|c|}
		\hline
		Salt & $D^{+}$, $10^{-9}$ m$^2$/s & $D^{-}$, $10^{-9}$ m$^2$/s & $\beta $ & $\mathrm{Pe}$ \\ \hline
		LiCl & 1.029 & 2.032 & -0.327 & 0.298 \\ \hline
		LiI & 1.029 & 2.045 & -0.330 & 0.297 \\ \hline
		LiNO$_3$ & 1.029 & 1.902 & -0.297 & 0.311 \\ \hline
		NaI & 1.334 & 2.045 & -0.212 & 0.270 \\ \hline
		NaCl & 1.334 & 2.032 & -0.208 & 0.272 \\ \hline
		KCl & 1.957 & 2.032 & -0.017 & 0.229 \\ \hline
		KI & 1.957 & 2.045 & -0.021 & 0.228 \\ \hline
		KBr & 1.957 & 2.010 & -0.013 & 0.300 \\ \hline
		\hline
		LiCH$_3$COO & 1.029 & 1.089 & -0.028 & 0.431 \\ \hline
		KCH$_3$COO & 1.957 & 1.089 & 0.286 & 0.299 \\ \hline
		NaCH$_3$COO & 1.334 & 1.089 & 0.100 & 0.377 \\ \hline
	\end{tabular}
\end{table}

In Table~\ref{table:beta} we present the values of $\beta$ for some standard univalent (inorganic and organic) salts, together with their P\'{e}clet numbers, both are calculated using data on diffusion coefficients by~\citet{vanysek.r:2000}. It can be seen that inorganic salts of Li provides the largest negative $\beta$, which is roughly $-0.3$ for all of them.


If we further assume that the slit is long, i.e. $H \ll L$, one can take advantage of the lubrication approximation. The latter  implies that the dimensional derivatives of the ion fluxes and the fluid velocity in Eqs.~\eqref{NPH} and \eqref{cont} in the $X-$direction are small compared to those in the $Z-$direction:%
\begin{equation}
\left\vert \partial _{X}J_{x}^{\pm }\right\vert \propto \frac{C^{\pm }U}{L}\ll
\left\vert \partial _{Z}J_{z}^{\pm }\right\vert \propto \frac{C^{\pm }U}{H},
\label{lub1}
\end{equation}%
\begin{equation}
\left\vert \partial _{X}U_{x}\right\vert \propto \frac{U}{L}\ll \left\vert
\partial _{Z}U_{z}\right\vert \propto \frac{U}{H},  \label{lub2}
\end{equation}%
where $U$ is the characteristic velocity.

\section{Electrostatic disturbance and surface potential}\label{sec:potential}

For a long channel the ion
concentrations obey local Boltzmann distributions at any cross-section \cite{fair1971,peters.pb:2016}:
\begin{equation}
c^{\pm }=c_m\left( x\right) \exp \left( \mp \phi \right) ,  \label{cpm}
\end{equation}%
where
\begin{equation}
\phi =\psi (x,z)-\psi_m\left( x\right),   \label{fd}
\end{equation}%
which is Eq.~\eqref{eq:PSI} rewritten in dimensionless form.
Here the midplane (``bulk'') concentration $c_m$ and the potential $\psi_m$
vary only in $x$ direction. The boundary conditions for $c_m$ are
\begin{equation}
c_m\left( 0\right)  =1,\quad c_m \left( 1\right) =c_{1}=C_{1}/C_{0}.  \label{c0}
\end{equation}

We also set
\begin{equation}
\psi_m\left( 0\right)  =0, \label{psi0}
\end{equation}%
but the value of $\psi_m\left( 1\right) $ has to be determined.

Since the leading-order term in the Poisson equation \eqref{PEq} involves only derivative with respect to $z$, the potential
$\phi $ at each cross-section satisfies the Poisson-Boltzmann equation
\begin{equation}
\partial _{zz}\psi =\partial _{zz}\phi =c_m\left( x\right) \lambda ^{-2}\sinh
\phi ,  \label{PB}
\end{equation}%
where
\[
\lambda = \frac{2 \lambda_{D}^{\star}}{H}.
\]%

To integrate Eq.~\eqref{PB} we should impose two electrostatic boundary conditions. Symmetry of the
channel dictates that
\[
\partial _{z}\psi \left( x,0\right) =\partial _{z}\phi \left( x,0\right) =0.
\]%
Another condition requires a constant surface charge density of the walls, which is equivalent to a constant gradient of the surface potential.
\begin{equation}
\partial_{z}\psi \left( x,1\right) =\partial _{z}\phi \left( x,1\right) =%
\frac{H}{\ell _{GC}}.  \label{cc}
\end{equation}%
However, the surface potential $\phi_s = \phi \left( x,1\right)$ varies with $x$ since a local Debye length scales with $c_m^{-1/2}.$ We emphasise that this potential is electrostatic  and in general differs from the so-called zeta potential $\zeta$, which is the measure of electro-osmotic mobility~\cite{vinogradova.oi:2023}. In the case of a thick slit obeying no-slip hydrodynamic boundary conditions, which we address here, $\zeta = \phi_s$, but for many surfaces and/or thin slits this is not so~\cite{joly.l:2004,silkina.ef:2020,asmolov.es:2025b}.

The first integration of the Poisson-Boltzmann equation \eqref{PB} leads to
the exact relation between the surface potential and charge~\cite{asmolov.es:2025}
\begin{equation}
\phi _{s} =  2\arsinh\left( \frac{\lambda _{D}^{\star}}{c_m^{1/2}\ell
_{GC}}\right),  \label{eq:pot-charge_hs}
\end{equation}
where the salt-dependent quantity $\lambda _{D}^{\star}/(c_m^{1/2}\ell_{GC})$ plays a role of the effective surface charge.

When the effective surface charge density  is small, Eq. (\ref{eq:pot-charge_hs}) reduces to
\begin{equation}
\phi_{s} \simeq \frac{2\lambda^{\star}_{D}}{c_m^{1/2}\ell _{GC}}.
\label{fs_DB}
\end{equation}%

If the effective surface charge is large, the surface potential depends on it weakly logarithmically:
\begin{equation}\label{fs_DB2}
\phi_{s} \simeq \pm 2 \ln \left[\dfrac{2 \lambda _{D}^{\star}}{c_m^{1/2}|\ell_{GC}|}\right]
\end{equation}
The choice of sign in \eqref{fs_DB2} depends on whether the walls are positively or negatively charged. The plus sign must be taken for a positive $\ell_{GC}$, and vice versa.

One further comment should be made. To incorporate a variable surface potential (i.e. a constant charge density)  some authors~\cite{kirby.bj:2004,ault.jt:2019,lee.s:2023} treated it for 1:1 salts as $\phi_s \simeq - a \ln c_m$, where $a$ is a fitting constant. To see that this approximation is unjustified, recast \eqref{fs_DB2} as
\begin{equation}\label{fs_DB2_note}
\phi_{s} \simeq \pm 2 \ln \left[\dfrac{2 \lambda _{D}^{\star}}{|\ell_{GC}|}\right] \mp \ln c_m.
\end{equation}
While Eq.~\eqref{fs_DB2_note} for highly charged surfaces indeed represents a linear relationship between $\phi_s$ and $\ln c_m$, it is obvious that the first term is too large to be omitted, and that $a=1$ for univalent salts. In the case of a weak effective surface charge the dependence of surface potential on concentration is not logarithmic. It follows from Eq.~\eqref{fs_DB} that  $\phi_s \propto c_m^{-1/2}$.

In any event, it seems clear that the variation in $\phi_s$ (of nonconducting materials) along the slit can be neglected only provided that, to first order, $c_m$ is constant. Such a situation naturally occurs when $c_m \simeq 1$, which is equivalent to $\Delta c \ll 1$, but if so, the diffusioosmosis would practically vanish. Note that \citet{ault.jt:2019} concluded that the constant potential approximation becomes justified, provided $\Delta c$ is small and/or both bulk solutions  are dilute. Our results do not support this claim, which has already been employed in some publications~\cite{lee.s:2023}. It follows from \eqref{eq:pot-charge_hs} that $\phi_s$ varies with $c_m$.  Hence, small, but realistic $\Delta c$ should be insufficient to neglect  the variation in $\phi$  even if both reservoirs are diluted [see also Appendix~\ref{a:1}]. Later we shall see that some portions of the slit could be of a constant $\phi_s$, but (i) not the whole slit and (ii) not necessarily at low midplane concentration.

Finally, performing the second integration in \eqref{PB} yields an exact analytical solution for the disturbance potential
 \begin{equation}\label{eq:PBSWexact}
 \phi = 4 \artanh \left[e^{(z-1)c_m^{1/2}/\lambda}\tanh \left(\dfrac{\phi_s}{4}\right)\right],
\end{equation}
where $\phi_s$ is given by \eqref{eq:pot-charge_hs}.


\section{Fluid flow rate}\label{sec:flowrate}

The flow rate $\mathcal{Q}$ of the fluid (expressed per unit film thickness),
\begin{equation}
\mathcal{Q}=\int_{0}^{1}u_{x}dz,  \label{jq}
\end{equation}%
is the same at any cross-section. From the solution of the Stokes equations \eqref{cont}, \eqref{mom} in the thick channel limit follows that the main contribution to $\mathcal{Q}$ is coming from its
central (``bulk'') part, where the fluid velocity is the sum of the plug diffusio-osmotic and parabolic pressure-driven
velocities:
\[
u_{x} \simeq u_{s}-\frac{1-z^{2}}{2}\partial
_{x}p,
\]%
where $u_s$ is the diffusioosmotic slip velocity, or, equivalently, the velocity of the diffusio-osmotic plug flow.
Consequently, the flow rate that obeys Eq.~\eqref{jq} is given by
\begin{equation}
\mathcal{Q}\simeq u_s-\frac{\partial _{x}p}{3},
\label{Q}
\end{equation}%
with the slip velocity given by~\cite{asmolov.es:2025}
\begin{equation}
u_{s}=-\frac{\partial _{x}c_{m}}{c_{m}} \left[\beta \phi _{s}+4\ln \left[ \cosh \left( \frac{\phi _{s}}{4}\right) \right]\right]. \label{ui2}
\end{equation}
Here the first and second terms are the electro- and chemi-osmotic contributions. This equation, which is identical to that derived by~\citet{prieve1984motion} for a single wall, can be recast as
\begin{equation}
u_{s}=\mathcal{F}\partial _{x}c_{m},  \label{uis}
\end{equation}
where
\begin{equation}
\mathcal{F}  = -\frac{ \beta \phi _{s}+4\ln \left[ \cosh \left( \phi _{s}/4\right) \right] }{c_{m}} = \mathcal{F}_e +  \mathcal{F}_c \label{uif}
\end{equation}
is  the function of $c_{m}$ solely [see Eq.~\eqref{eq:pot-charge_hs} for $\phi_s$].
The function $\mathcal{F}$ is the sum of electroosmotic $\mathcal{F}_e$ and chemiosmotic $\mathcal{F}_c $ contributions, which depend on
concentration differently. We also emphasise that a given $\mathcal{F}$ normally corresponds to two different
values of $\phi _{s}.$ It follows that $\phi _{s}$ cannot be  determined uniquely, if only $u_{s}$ of a given salt is known, and a supplementary information is required. We return to this point below.

In the case of low surface potentials, $\left\vert \phi _{s}\right\vert \leq
1$, one can use the approximation $\ln \left[ \cosh \left( \phi /4\right) %
\right] \simeq \phi ^{2}/32$. The chemiosmotic term in Eq.~\eqref{uif} then reduces to
\begin{equation}
\mathcal{F}_c\simeq -\frac{\phi _{s}^{2}}{8c_{m}}\ll 1  \label{mdo_sm}
\end{equation}%
Thus, this term turns out to be small compared to the
first one (i.e. electroosmotic) in Eq.~\eqref{uif} that is  of the order of $\phi _{s}$.

In the opposite case of $\left\vert \phi _{s}\right\vert \gg 1$ one
can approximate $\ln \left[ \cosh \left( \phi /4\right) \right] \simeq \left\vert \phi \right\vert
/4-\ln \left( 2\right)$
to obtain
\begin{equation}
\mathcal{F}_c\simeq -\frac{\left\vert \phi _{s}\right\vert }{c_{m}}\gg 1.  \label{mdo_la}
\end{equation}%
One can, therefore, conclude that for sufficiently large surface potentials chemiosmotic and electroosmotic
contributions are typically of the same order of magnitude.


Since $\mathcal{Q}$ is the same at any cross-section, it
may be determined by integrating Eq.~\eqref{Q} over $x$
\begin{equation}\label{eq:int_Q}
 \mathcal{Q} = \int_{0}^{1} \left[u_s - \frac{\partial _{x}p}{3} \right] dx.
\end{equation}
This can be re-written as
\begin{equation}
\mathcal{Q}=\int_{0}^{1}\mathcal{F}\partial _{x}c_{m}dx - \frac{\Delta p}{3}=\int_{1}^{c_{1}}%
\mathcal{F}dc_{m} - \frac{\Delta p}{3},  \label{qint}
\end{equation}%
where $\Delta p = p_1 - p_0$, and we conclude that compared to the  case of zero hydrostatic pressure drop the fluid flow rate is reduced by an amount $\Delta p/3$. In other words, $\mathcal{Q}$ is the sum of the integral diffusio-osmotic contribution and that of the pressure-driven flow. We remark and stress that the former does not change by applying an external pressure.

The reasons for differences in the measured $|\mathcal{Q}|$ for different ion pairs now becomes clear. For instance, $\beta$ of LiI is quite large and negative. It is smaller, but also negative for NaI, and KI has virtually zero $\beta$. It follows from  Eqs.~\eqref{uif} and \eqref{qint_l} that for negatively charged surfaces the magnitude of the diffusio-osmotic contribution to $\mathcal{Q}$ increases according to the sequence $\mathrm{KI}<\mathrm{NaI}<\mathrm{LiI}$. Such an order was observed by~\citet{lee.c:2014}  and indicates that their (silicon oxide) surfaces were negatively   charged. However, for positively charged walls a different story obtains. Based on Eq.~\eqref{uif} one can expect that the
highest $|\mathcal{Q}|$ given by \eqref{qint_l} will be observed for KI, the smaller for NaI, and so on. Thus, the uncertainty between two values of $\phi_s$ and hence $\ell_{GC}$ satisfying a measured value of $\mathcal{Q}$  can always be resolved by  performing measurements with several salts of different $\beta$.
\begin{figure}[h]
\begin{center}
\includegraphics[width=0.8\columnwidth]{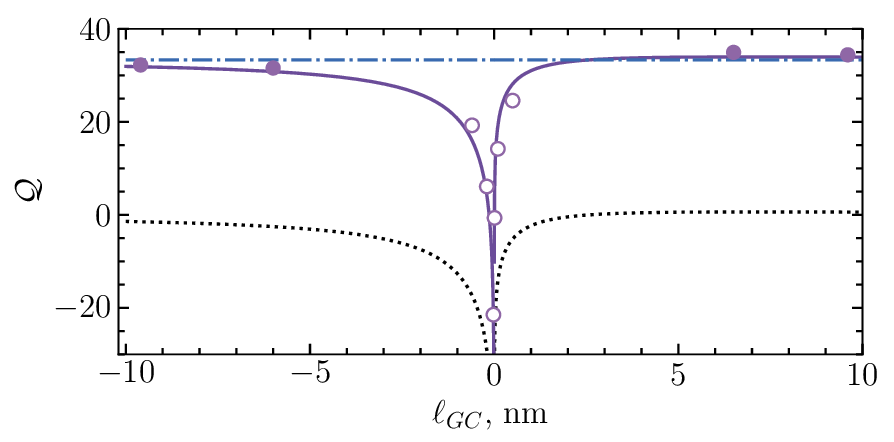}
\end{center}
\caption{$\mathcal{Q}$ as a function of $\ell_{GC}$ computed using $c_1 =
10^2$ and $\beta = -0.3$ for pressure drop $\Delta p = -10^{2}$ (solid curves) and $0$ (dotted curves). The dash-dotted curve corresponds to the case of $c_1 = 1$ and $\Delta p = -10^{2}$. Open and filled circles show calculations from Eqs.~\eqref{qint_l} and \eqref{eq:q_lin}.}
\label{fig:qb}
\end{figure}

By varying $\ell_{GC}$ it is possible to obtain the curves for $\mathcal{Q}$ shown in Fig.~\ref{fig:qb}.
To obtain $\mathcal{Q}$ at a given Gouy-Chapman length we first calculate $\mathcal{F}$ from Eq.~\eqref{uif} with $\phi_{s}$ related to $\ell_{GC}$ by \eqref{eq:pot-charge_hs} and then perform the numerical integration of Eq.~\eqref{qint}.
The calculations are made using $c_1 = 10^2$, and for this specimen example we employ typical for inorganic salts of lithium $\beta = -0.3$ (see Table~\ref{table:beta}). The results refer to fixed $\Delta p = 0$ (the dotted curve) and $-10^2$ (the solid curve). We begin with the lower (dotted) curve that corresponds to the case of the same hydrostatic pressure in both reservoirs. For negatively charged surfaces (positive $\beta \phi_s$) $\mathcal{Q}$ is always negative [since $\mathcal{F}$ given by \eqref{uif} is negative]. The fluid flow rate decreases monotonically from  zero at $\ell_{GC} \to - \infty$ down to $-\infty$ when the Gouy-Chapman length turns to zero.
Since for positively charged surfaces the numerator in \eqref{uif} can take any sign depending on
the value of $\phi_s$, the flow rate could be either positive or negative (or zero). On increasing $\ell_{GC}$ from zero (negative $\beta \phi_s$) the flow rate first increases sharply (from $-\infty$), then becomes positive but small in magnitude, and after exhibiting a maximum decreases slowly to zero. Summarizing, the net diffusio-osmotic flow is normally toward the ``fresh'' reservoir, and could be extremely large provided surfaces are highly charged. The flow toward the ``salty'' end emerges only if $\beta$ and $\phi_s$ are of the opposite sign and  surfaces are less charged. It can be seen that this (positive) flow is bounded and much smaller in magnitude compared to an attainable negative one. However, by applying external hydrostatic pressure to a low salinity reservoir  (which is equivalent to imposing a negative hydrostatic pressure drop/gradient between the reservoirs), we induce a supplementary Poiseuille flow toward the high concentration side, so the flow rate vs. $\ell_{GC}$ curve in Fig.~\ref{fig:qb} shifts toward a positive $\mathcal{Q}$.  We emphasize that to reverse the diffusio-osmotic flow quite small pressure drop is required - the need to counterbalance the huge osmotic pressure $\Delta \Pi \simeq 50$ bar as in reverse osmosis (exploiting semipermeable membranes) is thereby removed. Indeed, $\Delta p = -10^2$ used here corresponds to  $\Delta P \simeq -0.2$ bar only [for $H = 10^2$ nm], which can be easily achieved by using standard syringe pumps or simply by increasing the height of one of the reservoirs.
Also included in Fig.~\ref{fig:qb} the flow rate $\mathcal{Q}=-\Delta p/3 \simeq 33.3$ expected under the same applied pressure, but the vanishing of the concentration difference.

The integral in \eqref{qint} becomes divergent if $|\ell_{GC}| \to 0$. This singularity can be identified by Taylor expanding the expression for $u_s$ and $\phi_s$ at large $\lambda _{D}/\left\vert \ell _{GC}\right\vert$, which implies that $c_m \ll (\lambda _{D}^{\star }/\ell _{GC})^2$. Using \eqref{mdo_la} and \eqref{ui2} we find
\begin{equation}
u_{s} \simeq -\frac{\beta \phi _{s}+\left\vert \phi _{s}\right\vert }{c_{m}}%
\partial _{x}c_{m}
\end{equation}%
where $\phi _{s}$ is given by \eqref{fs_DB2}. Then by
integrating \eqref{qint} we obtain
\begin{equation}
\mathcal{Q}\simeq \left( \mp \beta -1\right) \ln
c_{1} \left( 2\ln \left( \frac{%
2\lambda _{D}^{\star }}{\left\vert \ell _{GC}\right\vert }\right) -\ln c_{1}\right)- \frac{\Delta p}{3}.  \label{qint_l}
\end{equation}
Here the upper (lower) sign corresponds to positive (negative) surface charge.
Note that Eq.~\eqref{qint_l} predicts a dependency of the flow rate on $\ln(c_1)$. For small $\Delta c$, however, it reduces to
\begin{equation}
\mathcal{Q}\simeq 2 \left( \mp \beta -1\right) \ln \left( \frac{%
2\lambda _{D}^{\star }}{\left\vert \ell _{GC}\right\vert }\right)\Delta c - \frac{\Delta p}{3},  \label{qint_l1}
\end{equation}
i.e. $\mathcal{Q}$   depends linearly on $\Delta c$.

The calculations from Eq.~\eqref{qint_l} are compared with the results for $\mathcal{Q}$ presented in Fig.~\ref{fig:qb} (the upper curve). It can be seen that \eqref{qint_l} provides a reasonably good fit to the numerical curve  up to $\left\vert \ell _{GC}\right\vert \simeq 1$ nm. At larger $\left\vert \ell _{GC}\right\vert $ the surface potential becomes smaller, so this simple analytical equation does not apply.

In the case of low surface potential [attained when $c_m \gg (\lambda _{D}^{\star }/\ell _{GC})^2$] the slip velocity can be approximated by~\cite{asmolov.es:2025}
\begin{equation}
u_{s}\simeq -2\beta \frac{\lambda _{D}^{\star }\partial _{x}c_{m}}{%
c_{m}^{3/2}\ell _{GC}}.  \label{u_DB}
\end{equation}%
By substituting this equation into (\ref{qint}) we obtain
\begin{equation}
\mathcal{Q} \simeq -4\beta \frac{\lambda _{D}^{\star }}{\ell _{GC}}\left(
1-c_{1}^{-1/2}\right) - \frac{\Delta p}{3}.  \label{eq:q_lin}
\end{equation}%
The calculations from \eqref{eq:q_lin} are also included in Fig.~\ref{fig:qb}. The approximations are quite good for large $|\ell _{GC}|$ (small $|\phi_s|$). We emphasize that Eq.~\eqref{eq:q_lin} will fail at moderate $\ell _{GC}$. For example, it will not describe the maximum and sign reversal of $\mathcal{Q}$.
Note that Eq.~\eqref{eq:q_lin} also accounts for specific ion selectivity and is linear in $\Delta c$, when the latter is small
\begin{equation}
\mathcal{Q} \simeq 2\beta \frac{\lambda _{D}^{\star }}{\ell _{GC}} \Delta c - \frac{\Delta p}{3}.  \label{eq:q_lin2}
\end{equation}%

\begin{figure}[h]
	\begin{center}
		\includegraphics[width=0.8\columnwidth]{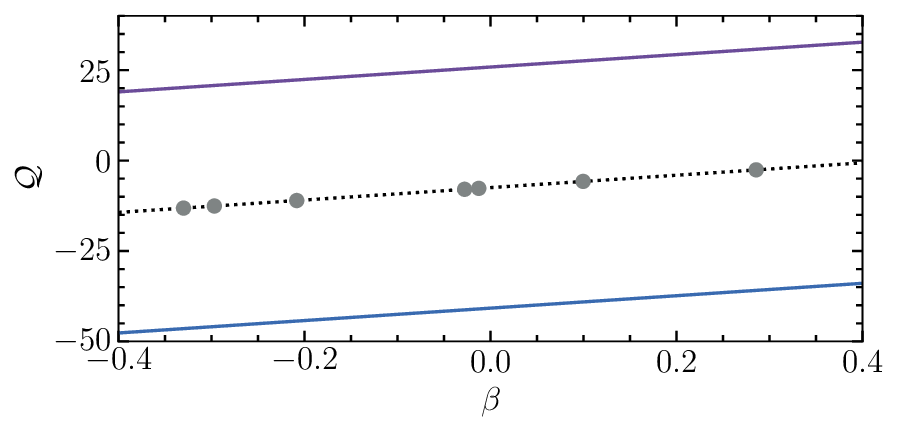}
	\end{center}
	\caption{The flow rate $\mathcal{Q}$ as a function of $\beta$ computed using $\ell_{GC} = -1$ nm for $c_1 = 10^2$, $\Delta p = -10^{2}$, 0 and $10^{2}$ (from top to bottom). Circles from left to right mark values
		of $\mathcal{Q}$ that correspond to LiI, LiNO$_3$, NaCl, LiCH$_3$COO, KBr,
		NaCH$_3$COO, and KCH$_3$COO.  }
	\label{fig:q_beta}
\end{figure}

The results displayed in Fig.~\ref{fig:qb} refer to fixed $\beta = -0.3$ typical for inorganic lithium salts. It is also of interest to consider the situations of $\beta = 0$ (inorganic potassium salts, lithium acetate) and $\beta = 0.3$ (potassium acetate). In other calculations (not shown) we used these two values of $\beta$. When $\beta = 0$ electroosmosis does not emerge. In this case left and right branches of the flow rate curves are symmetric and do not exhibit a maximum. Without applied hydrostatic pressure $\mathcal{Q}$ takes only negative values, but, of course, it can become positive if a  negative $\Delta p$ is applied. With $\beta = 0.3$ the plot represents a mirror image of Fig.~\ref{fig:qb} and the maximum of $\mathcal{Q}$ is observed for negatively charged surfaces. This effect has been noted
before~\cite{asmolov.es:2025}.

If we keep $\ell_{GC}$  fixed, but vary $\beta$, we move to the situation displayed in Fig.~\ref{fig:q_beta}.
The calculations are made using $\ell_{GC} = -1$ nm and the same value of $\mathcal{C}_0$ and $c_1$ as in Fig.~\ref{fig:qb}.
It can be seen that the flow rate increases linearly with $\beta$. Figure~\ref{fig:q_beta} is also intended to illustrate that depending on $\beta$ and $\Delta p$, the flow rate can be either positive or negative. We have marked with circles the values of $\mathcal{Q}$ obtained at $\Delta p = 0$ that correspond to different salts from Table~\ref{table:beta}.  For all of them $\mathcal{Q}$ is negative, but different in magnitude. It is evident that to attain the same flow rate of, say, LiI and KCH$_3$COO solutions some negative pressure drop would be necessary for the former (or a positive one to the latter). If we make similar calculations with positively charged surfaces, $\mathcal{Q}$ will decrease with $\beta$, i.e. is largest for LiI and smallest for  KCH$_3$COO.

Taken together, the results suggest that the fluid flow rate $\mathcal{Q}$ can be tuned by applying appropriate $\Delta p$. However, two distinct $\ell_{GC}$ would lead to the same flow rate, and this should be remembered when an experimentally measured $\mathcal{Q}$ is used to infer the surface potential or charge density~\cite{ault.jt:2019,lee.c:2014}. If, say, $\mathcal{Q} = 50$ is measured for $c_1 = 10^2$ by applying $\Delta p = -160$,  this corresponds to both $\ell_{GC} = 0.8$ and $-4.6$ nm.

 Below we shall see that concentration and potential profiles can be expressed in terms of $\mathrm{Pe }\mathcal{Q}$. Thus, by tuning $\Delta p$ one can easily adjust their form for relevant applications.

\section{Ion fluxes and concentrations}\label{sec:fluxes2}

We introduce the ion fluxes $\mathcal{J}^{\pm }$ as
\begin{equation}
\mathcal{J}^{\pm }= \int_{0}^{1}j_{x}^{\pm }dz,  \label{jq1}
\end{equation}%
and recall that they are constant along the channel as the flow rate is. If we make the additional assumption that the main contributions to
$\mathcal{J}^{\pm }$ come from the central (``bulk'') part, where the ion concentrations are close to $c_m$,  integrating Eq.~(\ref{NP1}) over $z$ one obtains
\begin{equation}
\mathcal{J}^{+}\simeq c_{m}\mathrm{Pe}\mathcal{Q}+\left( 1+\beta \right)
\left( -\partial _{x}c_{m}-c_{m}\partial _{x}\psi _{m}\right) ,  \label{jp}
\end{equation}%
\begin{equation}
\mathcal{J}^{-}\simeq c_{m}\mathrm{Pe}\mathcal{Q}+\left( 1-\beta \right)
\left( -\partial _{x}c_{m}+c_{m}\partial _{x}\psi _{m}\right) .  \label{jm}
\end{equation}
These equations, which form the basis of our analysis below, do not apply once the EDL contribution to $\mathcal{J}^{\pm }$ (or conductivity
) dominates over the bulk one. The limit of their validity can be evaluated in terms of the Dukhin number, $\mathrm{Du} \leq 1$, usually introduced
for colloid particles~\cite{dukhin.ss:1993}. Following~\citet{bocquet.l:2010} we define the upper value of the Dukhin length (at the ``fresh'' end):
\begin{equation}
	\ell^{\star}_{Du} = \frac{\lambda_D^{\star 2}}{|\ell_{GC}|}.
\end{equation}
Note that a useful formula for univalent electrolyte would be $\ell^{\star}_{Du} \simeq 2.5 \times 10^{-3} \dfrac{\sigma [\mathrm{mC/m^2}]}{\mathcal
{C}_0 [\mathrm{mol/l}]}$. Taking then into account the recent conductivity analysis~\cite{mouterde.t:2018,vinogradova.oi:2021}, we might argue that
for a thick slit $\mathrm{Du}  \simeq 4\ell^{\star}_{Du}/H $. It follows that \eqref{jp} and \eqref{jm} are sensible approximations when $H \geq 4 \ell^{\star}_{Du}$. To put in numbers,
for $\mathcal{C}_0 = 10^{-3}$ mol/l and $\ell_{GC} = 1$ nm the slit thickness should exceed 400 nm.


Introducing $\mathcal{J} = \mathcal{J}^{\pm}$, one can exclude $\mathcal{Q}$ by
subtracting these equations to get
\begin{equation}
\partial _{x}\psi _{m}=-\beta \frac{\partial _{x}c_{m}}{c_{m}}.  \label{pot2}
\end{equation}
Integrating Eq.~(\ref{pot2}) and imposing condition (\ref{c0}) yields
\begin{equation}
\psi _{m}=-\beta \ln c_{m}.  \label{eq:phi0_out}
\end{equation}%

Substituting (\ref{pot2}) into
(\ref{jp}) we obtain an ordinary differential equation
\begin{equation}
\mathcal{J}=c_{m}\mathrm{Pe}\mathcal{Q}-\left( 1-\beta ^{2}\right) \partial
_{x}c_{m}  \label{dc}
\end{equation}%
that provides a route to the determination of
$c_{m}\left( x\right)$. The last result also follows directly from Eqs.~\eqref{jp} and \eqref{jm}.

By integrating Eq.~\eqref{dc} and applying the first boundary condition in \eqref{c0} we derive
\begin{equation}
c_{m}=\left( 1-\frac{\mathcal{J}}{\mathrm{Pe}\mathcal{Q}}\right) \exp \left(
\frac{\mathrm{Pe}\mathcal{Q}}{1-\beta ^{2}}x\right) +\frac{\mathcal{J}}{%
\mathrm{Pe}\mathcal{Q}}.  \label{cx}
\end{equation}%
The form of this equation is identical to our earlier formula~\cite{asmolov.es:2025}, but now $\mathcal{Q}$ and $\mathcal{J}$ depend on $\Delta p$. We also emphasise that the lubrication approximation we employ here justifies even a strongly non-linear variation in local concentration
along $x$ and that its $x$-derivative can, in principle, be any. It is only necessary that the slit is long enough to ensure that the dimensional $X$-derivative (of the order of $C_{m}/L$), is much less than the $Z$-derivative (which is of the order of $C_{m}/H$).

The flux $\mathcal{J}$ can then be found from (\ref{cx}) by using the second boundary condition in \eqref{c0}
\begin{equation}\label{JQ}
\mathcal{J}=\mathrm{Pe}\mathcal{Q}\left[1 + \frac{\Delta c}{1-c^{\ast }}\right],
\end{equation}
where the (positive) quantity
\begin{equation}
c^{\ast }=\exp \left( \frac{\mathrm{Pe}\mathcal{Q}}{1-\beta ^{2}}\right)
\label{cc1}
\end{equation}
increases exponentially from practically zero (at large negative $\mathrm{Pe} \mathcal{Q}$ to $\infty$ when $\mathrm{Pe}\mathcal{Q} \to \infty$.

Equation (\ref{cx}) can be rewritten by using Eq.~\eqref{JQ} as
\begin{equation}
c_{m}= 1 + \frac{\Delta c}{c^{\ast }-1} \left[\exp \left( \frac{\mathrm{Pe}\mathcal{Q}}{%
1-\beta ^{2}}x\right) - 1\right].
\label{cx1}
\end{equation}
We thus excluded $\mathcal{J}$ from the expression for $c_{m}$. This result is crucial since clarifies that the local concentration profile along the slit for a specific salt (characterized by $\beta$ and $\mathrm{Pe}$) is determined only by the fluid flow rate $\mathcal{Q}$ and the concentration drop $\Delta c$ between reservoirs. It follows that the  flow rate at a given concentration drop, which is measurable~\cite{lee.c:2014}, can be used to uniquely infer the concentration distribution inside the slit that still remains ``invisible'' for modern experiment.

Summarizing, the set of Eqs.~\eqref{cx}-\eqref{cc1} we derived is the same as those  appropriate to the case of $\Delta p = 0$. What is significant is the novel expression for  $\mathcal{Q}$ [see \eqref{qint}], which now incorporates the pressure drop, as this can change not only the magnitude and sign of the flow rate, but those of $\mathcal{J}$, as well as the concentration and potential profiles. We return to these issues below.

\begin{figure}[h]
	\begin{center}
		\includegraphics[width=0.8\columnwidth]{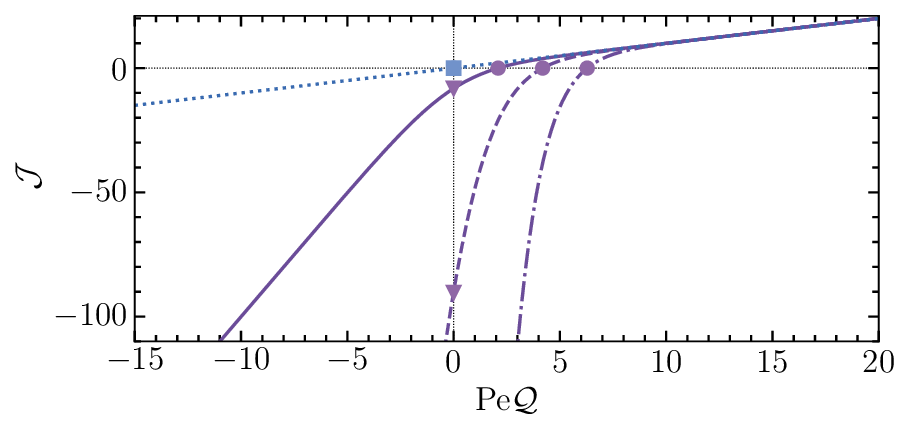}
	\end{center}
	\caption{$\mathcal{J}$ as a function of $\mathrm{Pe}\mathcal{Q}$ computed using $\beta = -0.3$ and $c_1 = 1$ (dotted), $10$ (solid), $10^{2}$ (dashed), $10^{3}$ (dash-dotted). The square marks the point of $\mathcal{J} = \mathcal{Q} = 0$. Circles and triangles indicate $\mathcal{J} = 0$ and $\mathcal{Q} = 0$.  }
	\label{fig:jb0}
\end{figure}

In Fig.~\ref{fig:jb0} we plot $\mathcal{J}$ as a function of $\mathrm{Pe} \mathcal{Q}$. The calculations are made from Eq.~\eqref{JQ} for several $c_1$ from 1 to $10^3$ using $\beta = -0.3$, typical for inorganic salts of lithium and potassium acetate. It can be seen that when $\mathrm{Pe}\mathcal{Q} \gg 1$ all curves practically merge into a single straight line and ionic flux $\mathcal{J}$ is positive, i.e. toward the ``salty'' reservoir. At smaller $\mathrm{Pe} \mathcal{Q}$ the calculated for different $c_1$ curves do separate. For a given $\mathcal{J}$, $\mathrm{Pe} \mathcal{Q}$ becomes greater as $c_1$ augments. At sufficiently large negative $\mathrm{Pe} \mathcal{Q}$ the curves become linear again, and their slope now augments with $c_1$. We also note that in this mode $\mathcal{J}$ is negative. In Fig.~\ref{fig:jb0} we marked with the square  the point of $\mathcal{J} = \mathcal{Q} = 0$ on the line corresponding to $\Delta c = 0$, where diffusio-osmosis is not generated. The intersections of the other curves ($\Delta c \neq 0$) with the horizontal line $\mathcal{J}=0$ are marked with circles, and with the vertical line $\mathcal{Q} = 0$ are shown by triangles.

\begin{figure}[h]
	\begin{center}
		\includegraphics[width=0.8\columnwidth]{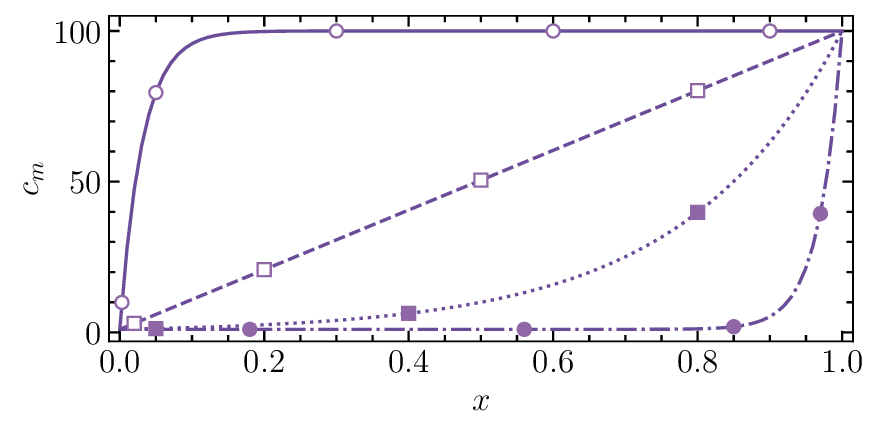}\\[0pt]
	\end{center}
	\caption{Concentration $c_m$ as a function of $x$
		for NaCl [$\mathrm{Pe} = 0.272$, $\beta = -0.208$]. Solid and dash-dotted curves show calculations using $\mathrm{Pe}\mathcal{Q} = -30$ and 30. Dashed and dotted curves correspond to the cases when $\mathcal{Q}=0$ and $\mathcal{J}=0$. Filled and open circles are obtained using Eqs.~\eqref{cx_largePeQ} and \eqref{cx_minus}. Filled and open squares show predictions of Eqs.~\eqref{cx_zeroJ} and \eqref{eq:c_lin2}.}
	\label{fig:c}
\end{figure}

It is of considerable interest to ascertain how the concentration profile along the slit depends on $\Delta p$, which impacts $\mathrm{Pe} \mathcal{Q}$.  Figure~\ref{fig:c} includes concentration curves for the special situations described above. The calculations are made from Eq.~\eqref{cx1} for NaCl [see its $\beta$ and $\mathrm{Pe}$ in Table~\ref{table:beta}] using $c_1 = 10^2$ and several $\Delta p$ that correspond to the cases of $\mathrm{Pe}\mathcal{Q} = -30$, $\mathcal{Q}=0$, $\mathcal{J}=0$, and $\mathrm{Pe}\mathcal{Q} = 30$ (from top to bottom). These parameters can only be set by applying an appropriate pressure drop. For example, if, say, we fix $\ell_{GC} = -1$ nm, which is close to the value inferred  from electrokinetic measurements in silica nanochannels~\cite{stein.d:2004}, the above situations would occur when using $\Delta p \simeq 3 \times 10^2$, $-33$, $-81$,  $-3.67 \times 10^2$, correspondingly. Note that for $\Delta p = 0$ the concentration distribution for this inorganic salt is convex at any $\ell_{GC}$ (not shown).  On applying positive $\Delta p$ the concentration curve becomes more convex near the ``fresh'' side [and simultaneously less convex near the ``salty'' one]. When the absolute value of negative $\mathrm{Pe}\mathcal{Q}$ is large enough, $c_m$ saturates to $c_1$ inside the slit. The uppermost curve in  Fig~\ref{fig:c} illustrates well that at $\mathrm{Pe}\mathcal{Q} = -30$ the  ``bulk''  concentration is constant and equal to that in the high salinity reservoir in the most of the slit. However, a negative hydrostatic pressure drop first makes the concentration profile less convex and then rectifies it [when $\mathcal{Q}=0$ is reached]. If we decrease (i.e. increase in magnitude) the negative pressure drop, the profile becomes concave. On decreasing $\Delta p$ further we move to the situation described by the lowermost curve in  Fig~\ref{fig:c}. Namely, in the most of the slit the ``bulk'' concentration is constant and equal to $c_0$, its rapid increase to $c_1$ is observed only in the vicinity of the ``salty'' reservoir.
Below we examine these cases further.


\subsection{Large positive $\mathrm{Pe} \mathcal{Q}$.}

From Eq.~\eqref{dc} it follows that if $c_0=c_1=1$ the concentration $c_m$ would also be equal to unity at any $x$, which yields $\mathcal{J}=\mathrm{Pe}\mathcal{Q}$. In this case diffusio-osmotic flow does not emerge and the fluid flow rate $\mathcal{Q}=-\Delta p/3$ could take on nonzero values only if $\Delta p \neq 0$, so is $\mathcal{J}$.  We emphasize that the pressure-driven flow transfers ions in the central electro-neutral region, so $\mathcal{J} \neq 0$, but since $\mathcal{J}^- = \mathcal{J}^+$ the electric current is not generated.
The collapse of  all curves into a straight line derived for  $\Delta c = 0$ at $\mathrm{Pe}\mathcal{Q} \gg 1$ follows from Eq.~\eqref{JQ}. Since
an increasing exponentially with  $\mathrm{Pe}\mathcal{Q}$ quantity $c^{\ast}$  becomes significantly greater than $c_1$, the second term in the brackets can be neglected.
The asymptotic behavior of $\mathcal{J}$ given by \eqref{JQ} is thus determined by $\mathcal{J}\simeq\mathrm{Pe}\mathcal{Q}$, and
Eq.~\eqref{cx1} reduces to a more compact expression
\begin{equation}
c_{m} \simeq  1 + \Delta c  \exp \left(
\frac{\mathrm{Pe}\mathcal{Q}}{1-\beta ^{2}}(x-1)\right).  \label{cx_largePeQ}
\end{equation}%
The calculations from Eq.~\eqref{cx_largePeQ} are compared with the results presented in Fig.~\ref{fig:c} (the lowermost curve). It can be seen that \eqref{cx_largePeQ} provides a very good fit to the curve calculated from exact Eq.~\eqref{cx1}.

\begin{figure}[h]
	\begin{center}
		\includegraphics[width=0.8\columnwidth]{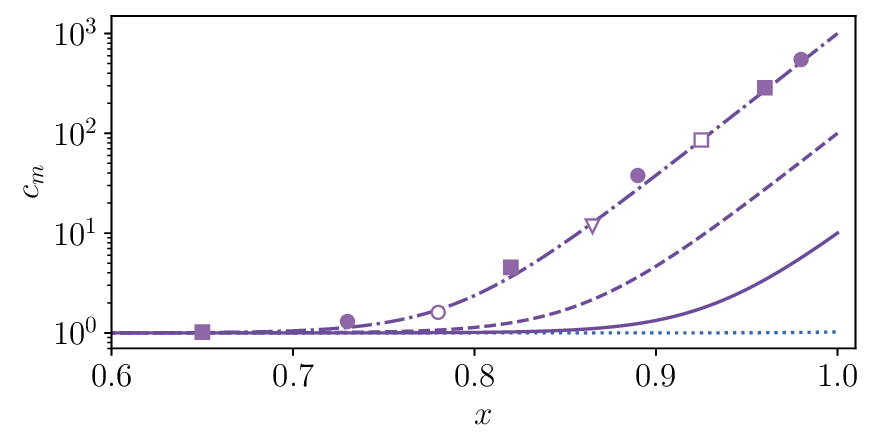}
	\end{center}
	\caption{Concentration profiles  calculated using $\mathrm{Pe} \mathcal{Q} = 30$ for $\beta = -0.3$ and $c_1 = 1$ (dotted), $10$ (solid), $10^{2}$ (dashed), $10^{3}$ (dash-dotted). The filled circles and squares are calculations for KCl and NaCl. The open circle, triangle, and square show $c_m$ for LiI, LiCl, and LiNO$_3$.   }
	\label{fig:conc_inf}
\end{figure}

The midplane concentration profiles of electrolyte then have the form shown (in a log-linear scale) in
Fig.~\ref{fig:conc_inf}. The calculations are made from Eq.~\eqref{cx1} using the same values of $\beta$ and $c_1$ as in Fig.~\ref{fig:jb0}. Also included is the concentration profile calculated with $c_1 = 10^3$. We also set $\mathrm{Pe}
\mathcal{Q} = 30$ by setting $\mathrm{Pe} = 0.3$ [lithium inorganic salts and potassium acetate] and $\mathcal{Q} = 100$. We recall that such a
value of $\mathcal{Q}$ cannot be attained without applied negative pressure or with moderate $\Delta p$ used in Fig.~\ref{fig:qb}. However, one can
easily provide $\mathcal{Q} = 100$ by applying a larger pressure drop. Say, for surfaces of $\ell_{GC} = -1$ nm, $\Delta p \simeq -3.38 \times 10^2$ is required.
It can be seen that in this mode all concentration profiles are almost constant, $c_m \simeq c_0 \simeq 1$, throughout most of the slit, only near the ``salty'' end they vary by exhibiting a rapid exponential increase of $c_m$ to $c_1$.
If we make another calculation in which $\beta = 0$ is set, this has the effect of slightly higher $c_m$  (in the range of $x$ from ca. 0.7 to 0.95) than obtained for $\beta=-0.3$. The discrepancy is hardly discernible at a moderate $c_1$, but augments on increasing it. Instead of plotting these curves of zero $\beta$ we include in Fig.~\ref{fig:conc_inf} the results of calculations from~\eqref{cx1} made for KCl that is having a practically vanishing $\beta$ (see Table~\ref{table:beta}). They are shown only for  $c_1 = 10^3$, where the discrepancy is largest. Also included in Fig.~\ref{fig:conc_inf} are calculations for inorganic lithium salts of $\beta \simeq -0.3$, which practically coincide with the (upper) curve obtained using $\beta \simeq -0.3$. Finally, we include the results obtained for NaCl ($\beta \simeq -0.208$) and conclude that they are confined between lithium salts and KCl.

\subsection{Zero $\mathcal{J}$.}

The intersections of the horizontal line $\mathcal{J} = 0$ with the computed $\mathcal{J}$-curves determine this $\mathcal{Q}$ of zero ionic flux. In the absence of concentration drop it is equal to zero, but is always positive for a positive  $\Delta c$ as
follows from Fig.~\ref{fig:jb0}. We emphasise that the positive global flow rate implies that to lock the channel for ion fluxes $\Delta p$ should be negative.

It follows from \eqref{JQ} that the ion flux disappears when
\begin{equation}\label{eq:zeroJ1_cm}
\mathrm{Pe}\mathcal{Q}\frac{c_{1}-c^{\ast }}{1-c^{\ast }} = 0,
\end{equation}
where $c^{\ast }$ is given by \eqref{cc1}. The strategy for suppressing the ion flux is apparent now. Suppose $c_1$ is set. The flow rate should then be adjusted (by pressure) to provide
\begin{equation}\label{eq:zeroJ1}
 \mathrm{Pe } \mathcal{Q} = (1-\beta^2) \ln c_1 \geq 0.
\end{equation}
The calculations from Eq.~\eqref{eq:zeroJ1} are included in Fig.~\ref{fig:jb0} to mark the points of zero ionic flux at different concentration drops.

The corresponding concentration profile is obtained by substituting \eqref{eq:zeroJ1} in \eqref{cx} and prescribing zero $\mathcal{J}$
\begin{equation}
c_{m}= \exp \left(x \ln c_1\right).  \label{cx_zeroJ}
\end{equation}%
From this equation it follows that the profile of concentration is convex. The calculations from Eq.~\eqref{cx_zeroJ} are included in Fig.~\ref{fig:c}. It can be seen that the agreement with the curve calculated from \eqref{cx1} is excellent.

Alternatively to provide $\mathcal{J} = 0$ at a prescribed $\mathcal{Q} \neq 0$  one can tune $c_1$:
\begin{equation}\label{eq:zeroJ2}
 c_1 =  \mathrm{Pe } \mathcal{Q} \exp (1-\beta^2).
\end{equation}

\subsection{Zero $\mathcal{Q}$.}

Another case of special interest is that for which the  fluid flow rate vanishes.
By expanding \eqref{JQ} about $\mathcal{Q}=0$ and taking the leading order term only yields
\begin{equation}\label{eq:zeroQ}
 \mathcal{J} \simeq -\left( 1-\beta ^{2}\right) \Delta c \leq 0.
\end{equation}
This result can also be obtained directly from the solution to Eq.~\eqref{dc} satisfying \eqref{c0}. Equation~\eqref{eq:zeroQ} points out that in the limiting case of equal concentrations in both reservoirs ($c_1 = 1$ or, equally, $\Delta c = 0$) the ion flux disappears, i.e. $\mathcal{J} = 0$. This is the only situation when the system is at equilibrium. If simultaneously $\Delta c \neq 0$ and $\mathcal{Q}$ vanishes, the ion flux
$\mathcal{J}$ takes on nonzero (negative) values, i.e. ions always propel toward the ``fresh'' bath.
The local ``bulk'' concentration may be determined from Eq.~\eqref{dc}, which now reduces to $\mathcal{J}=-\left( 1-\beta ^{2}\right) \partial
_{x}c_{m}$. Performing the integration, imposing conditions \eqref{c0}, and making use of \eqref{eq:zeroQ} we find
\begin{equation}
c_{m} \simeq 1-\frac{\mathcal{J}}{1-\beta ^{2}}x = 1+x\Delta c,
\label{eq:c_lin2}
\end{equation}%
This expression provides an excellent fit of the straight line in Fig.~\ref{fig:c}.

Thus in the absence of global fluid flow the concentration $c_m$ augments linearly with $x$. This implies that (in general) nonlinear concentration profiles can always be rectified (converted to linear) by completely suppressing $\mathcal{Q}$, e.g. by applied pressure.

%

\subsection{Large negative $\mathrm{Pe} \mathcal{Q}$.}

Finally, we focus on the case of a large negative $\mathrm{Pe} \mathcal{Q}$. If so,  the concentration rapidly increases from $c_0$ to $c_1$ near a ``fresh'' reservoir and becomes a constant in the rest of the slit (see, for example, the uppermost curve in Fig.~\ref{fig:c} for NaCl obtained using $c_1 = 10^2$). Under these conditions a limit $\mathrm{Pe} \mathcal{Q} \to - \infty$ can be applied that yields $\mathcal{J} \simeq \mathrm{Pe} \mathcal{Q} c_1$ and
\begin{equation}
c_{m} \simeq -\Delta c \exp  \left( \frac{\mathrm{Pe}\mathcal{Q}}{1-\beta ^{2}}x\right) + c_1.  \label{cx_minus}
\end{equation}%
The calculation from Eq.~\eqref{cx_minus} made for $c_1=10^2$ are included in Fig.~\ref{fig:c} and fully coincide with the corresponding exact curve.

\begin{figure}[h]
	\begin{center}
		\includegraphics[width=0.8\columnwidth]{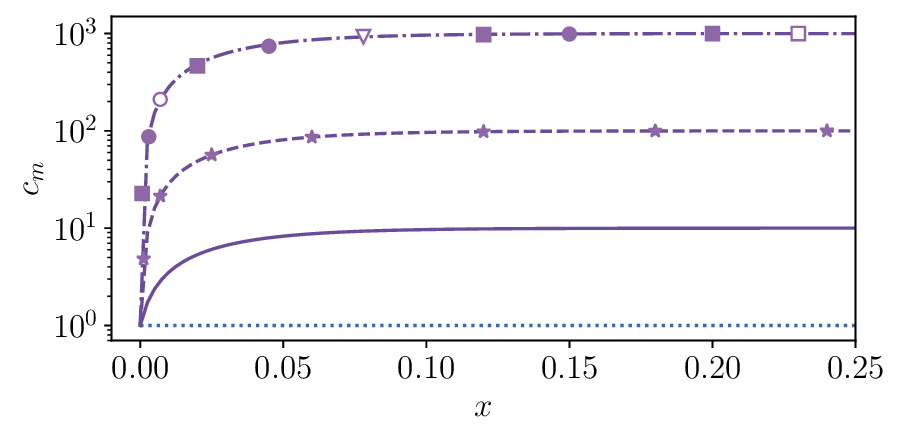}
	\end{center}
	\caption{Concentration profiles  calculated using $\mathrm{Pe} \mathcal{Q} = -30$ for $\beta = -0.3$ and $c_1 = 1$ (dotted), $10$ (solid), $10^{2}$ (dashed), $10^{3}$ (dash-dotted). Stars show calculations from Eq.~\eqref{cx_minus}. }
	\label{fig:conc_inf_min}
\end{figure}

To examine the significance of $\Delta c$ more closely, the concentration profiles at the short-distance region from the ``fresh'' reservoir are shown in  Fig.~\ref{fig:conc_inf_min}. The calculations are made from Eq.~\eqref{cx1} with the same parameters as in Fig.~\ref{fig:conc_inf}, but we now set $\mathrm{Pe} \mathcal{Q} = -30$. It can be seen that for all $c_1$ the curves saturate to $c_m = c_1$ very fast. The concentration plateau is reached already at $x \simeq 0.1$ or so.
Also included in Fig.~\ref{fig:conc_inf_min} are calculations  made with $c_1 = 10^3$ for several salts, the same ones as in Fig.~\ref{fig:conc_inf}. Finally, to illustrate the accuracy of~\eqref{cx_minus}, we include the calculations from this compact equation made using $c_1 = 10^2$.

\section{Midplane vs. surface potential}\label{sec:midpotential}

In our previous work~\cite{asmolov.es:2025} we have concluded that the diffusion (or liquid junction) potential $\Delta \psi_m = \psi_m(1)$ remains the same and cannot be tuned by the total flow rate (and hence $\Delta p$). However, the midplane potential $\psi_m$ given by Eq.~\eqref{eq:phi0_out} is determined by $c_m$ and hence is very sensitive to $\mathcal{Q}$ that, in turn, depends on the hydrostatic pressure drop as discussed above.

\begin{figure}[h]
	\begin{center}
		\includegraphics[width=0.8\columnwidth]{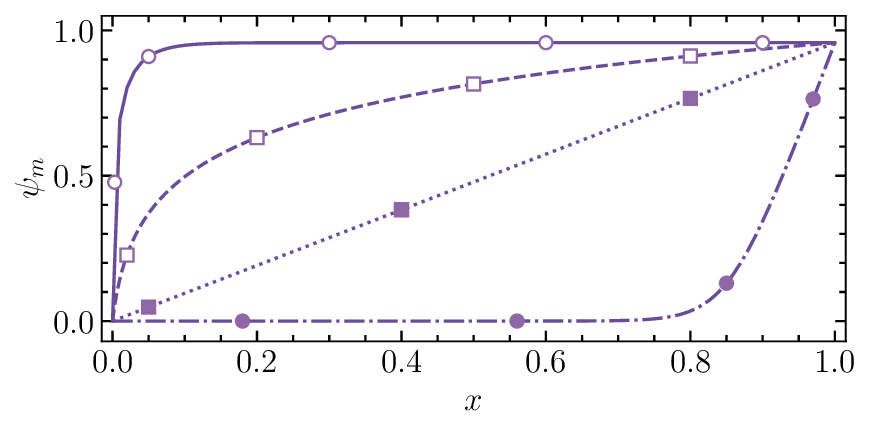}
	\end{center}
	\caption{Midplane potential $\psi_m$ vs $x$ computed from Eq.~\eqref{eq:phi0_out} for $c_m$ displayed in Fig.~\ref{fig:c}.  Filled and open circles are obtained using Eqs.~\eqref{psim_infQ} and \eqref{psim_infminQ}. Filled and open squares show predictions of Eqs.~\eqref{psim_J0} and \eqref{psim_Q0}.}
	\label{fig:psim}
\end{figure}

The profiles of $\psi_m (x)$ for special cases of large positive and negative $\mathrm{Pe} \mathcal{Q}$, which obviously represent
bounds on the attainable local midplane potentials, are given in Fig.~\ref{fig:psim}. The calculations are made from Eq.~\eqref{eq:phi0_out} using the concentration profiles of NaCl obtained from Eq.~\eqref{cx1}. Also included are calculations for the situations of zero $\mathcal{Q}$ and $\mathcal{J}$.

The lowermost curve corresponds to the limit $\mathrm{Pe} \mathcal{Q} \gg 1$, there $c_m$ is given by \eqref{cx_largePeQ}. For this special case Eq.~\eqref{eq:phi0_out} can be transformed to
\begin{equation}\label{psim_infQ}
\psi _{m} = -\beta \ln \left[1 + \Delta c \exp \left(
\frac{\mathrm{Pe}\mathcal{Q}}{1-\beta ^{2}}(x - 1) \right)\right].
\end{equation}

In the case of zero $\mathcal{J}$ the concentration $c_m$ is described by Eq.~\eqref{cx_zeroJ}. Substituting this to \eqref{eq:phi0_out}, we obtain
\begin{equation}\label{psim_J0}
\psi _{m} = -\beta x \ln c_1.
\end{equation}%
Thus, the $\psi _{m}$-profile becomes rectified, i.e. linear in $x$, if the ionic flux through the slit is blocked.
Then, performing similar calculations for the case of $\mathcal{Q}=0$, where the linear concentration profile is given by \eqref{eq:c_lin2}, we derive
\begin{equation}\label{psim_Q0}
\psi _{m} = -\beta \ln(1+x\Delta c)
\end{equation}%
and conclude that the potential in the central electro-neutral region depends on $x$ weakly logarithmically.

Finally, we turn to the case of large negative $\mathrm{Pe} \mathcal{Q}$ [the uppermost curve in Fig.~\ref{fig:psim}]. Substituting $c_m$
given by \eqref{cx_minus}
into \eqref{eq:phi0_out}
yields
\begin{equation}\label{psim_infminQ}
\psi _{m} = -\beta \ln \left[-\Delta c \exp  \left( \frac{\mathrm{Pe}\mathcal{Q}}{1-\beta ^{2}}x\right) + c_1\right].
\end{equation}%

The calculations from Eqs.~\eqref{psim_infQ}-\eqref{psim_infminQ} are included (symbols) in Fig.~\ref{fig:psim} and compared with prior calculations (curves) in which Eq.~\eqref{cx1} for the concentration profile has been employed. It can be seen that simple asymptotic equations provide an excellent fit to the exact analytical results.

In summary, although the diffusion potential is unaffected by $\Delta p$, the $\psi_m$-profiles inside the slit can be easily tuned by the pressure drop. This opens a possibility to tune the electric field (i.e. the potential gradient taken with minus) within the slit by just an external pressure, without using electrodes and a potentiostat.

Recall that the midplane concentration defines a local surface potential $\phi_s$ at a given surface charge density [see Eq.~\eqref{eq:pot-charge_hs}]. In turn, $\phi_s$ is a very important consideration in determining the diffusio-osmotic slip velocity $u_s$. Thus, before describing the results of calculations of $u_s$, it is instructive to consider the surface potential variations along the slit.

\begin{figure}[h]
	\begin{center}
		\includegraphics[width=0.8\columnwidth]{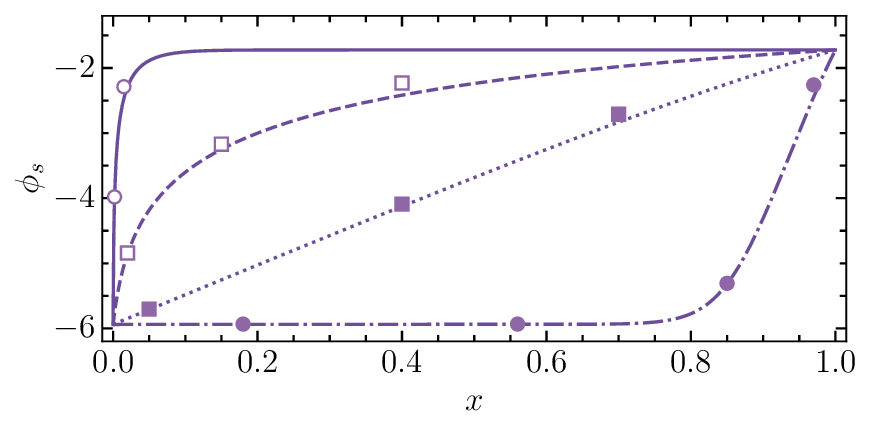}
	\end{center}
	\caption{Surface potential $\phi_s$ vs $x$ computed from Eq.~\eqref{eq:pot-charge_hs} using $\ell_{GC} = -1$ nm for $c_m$ displayed in Fig.~\ref{fig:c}. Filled and open circles show predictions of Eq.~\eqref{fs_DB2_note2} with $\psi_m$ calculated from Eqs.~\eqref{psim_infQ} and \eqref{psim_infminQ}. Filled and open squares are obtained using Eq.~\eqref{fs_DB2_note2} with $\psi_m$ given by Eqs.~\eqref{psim_J0} and \eqref{psim_Q0}. }
	\label{fig:phis}
\end{figure}

Figure~\ref{fig:phis} shows $\phi_s (x)$ for NaCl solution obtained from Eq.~\eqref{eq:pot-charge_hs} using $\ell_{GC} = -1$ nm, which gives $\lambda^{\star}/\ell_{GC} \simeq 10$, and provide concentration profiles shown in Fig.~\ref{fig:c}. For all curves $\phi_s$ varies from ca. $-6.0$ at $x=0$ to $-1.7$ at $x=1$, but the surface potential profiles along the slit are very different. With given parameters they (qualitatively) look similar to the profiles of $\psi_m$ displayed in Fig.~\ref{fig:psim}. For example, the $\phi_s$-curves obtained for large $|\mathrm{Pe} \mathcal{Q}|$ include the extended regions of constant surface potentials. However, the latter vary linearly with $x$, if $\mathcal{J} = 0$.

At first sight this is surprising, but the nature of such a qualitative similarity of $\psi_m$ and $\phi_s$ is apparent. Indeed, from Eqs.~\eqref{fs_DB2_note} and \eqref{eq:phi0_out} it follows that the surface potential for walls of a high  negative charge is related to $\psi_m$ linearly, as
\begin{equation}\label{fs_DB2_note2}
\phi_{s} \simeq  - 2 \ln \left[\dfrac{2 \lambda _{D}^{\star}}{|\ell_{GC}|}\right] - \dfrac{\psi_m}{\beta},
\end{equation}
where
\begin{equation*}
  - 2 \ln \left[\dfrac{2 \lambda _{D}^{\star}}{|\ell_{GC}|}\right] = \phi_s (0)
\end{equation*}
is the surface potential at $x=0$. Thus, the first term in \eqref{fs_DB2_note2} is a constant that depends on the bulk concentration in the ``fresh'' reservoir and surface charge density only, but does not depend on $x$. The second term is associated with the $x$-dependent contribution to $\phi_s$, since it is proportional to $\psi_m$, which varies from zero at $x=0$ to $-\beta \ln c_1$ at $x=1$. The surface potential drop throughout the slit is then given by $\Delta \phi_s = \phi_s (1)-\phi_s(0) = -\ln c_1$. Note that it neither depends on a specific $\phi_s-$profile, nor on $\beta$.

The calculations from Eq.~\eqref{fs_DB2_note2} for a corresponding special mode can be made
by inserting an appropriate  approximate expression for $\psi_m$ (see above). For example, using \eqref{psim_infQ} we can easily find the approximation for the surface potential in the limit of large $\mathrm{Pe} \mathcal{Q}$. Its asymptotic behavior in an adjacent to the ``salty'' reservoir region can be identified by Taylor  expanding about $x=1$. It
becomes immediately clear that,  to first order in $x$,
\begin{equation}
\phi_s \simeq \phi_s(0) + \ln c_1 + \frac{\Delta c}{c_1}\frac{\mathrm{Pe}\mathcal{Q}}{1-\beta ^{2}}(x - 1),
\end{equation}
i.e. the surface potential augments linearly with $x$.
Or, say, in the case of $\mathcal{J} = 0$ a substitution of \eqref{psim_J0} into Eq.~\eqref{fs_DB2_note2} yields a linear growth of $\phi_s$ in the whole slit, since
\begin{equation}
\phi_s \simeq \phi_s(0) +  x \ln c_1,
\end{equation}
and so on.

Such calculations are made and compared with the $\phi_s-$curves displayed in Fig.~\ref{fig:phis}.
It can be seen that Eq.~\eqref{fs_DB2_note2} gives an excellent match to the curves down to $|\phi_s| \simeq 3$, but at smaller $|\phi_s|$ there is
some discrepancy in the direction of smaller surface potential than predicted by using the exact concentration profiles (not shown). For example,  at $x = 1$ the exact $\phi_s \simeq -1.7$, but Eq.~\eqref{fs_DB2_note2} does predict $\simeq -1.3$, i.e. $24\%$ smaller in magnitude value. It must be remembered, however, that
this equation is derived  for large surface potentials only, and we did not expect it to be very accurate for the whole range of surface potentials in Fig.~\ref{fig:phis}.  

\section{Diffusio-osmotic slip velocity}\label{sec:slip_velocity}

We turn now to the apparent velocity of diffusio-osmotic slip, i.e. of the diffusio-osmotic plug flow in the electro-neutral part of the slit. It follows from Eq.~\eqref{ui2} that beside $\phi_s$ it is determined by $c_m$ and $\partial_x c_m$.

\begin{figure}[h]
	\begin{center}
		\includegraphics[width=0.8\columnwidth]{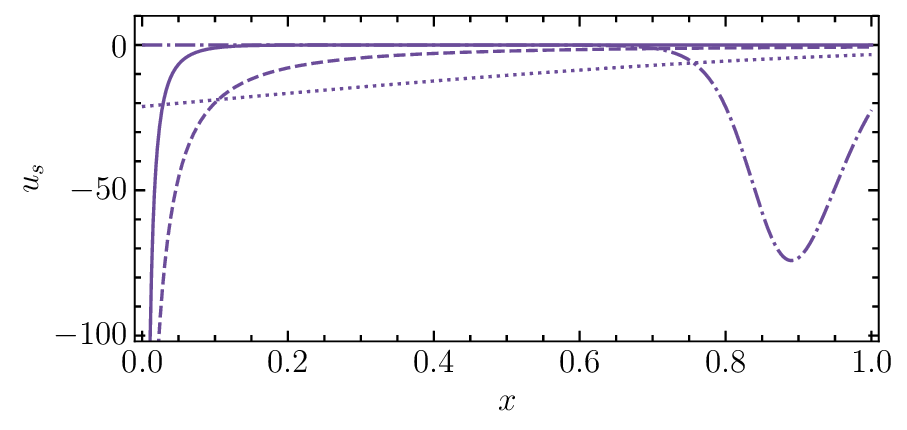}
	\end{center}
	\caption{Slip velocity as a function of $x$ for the same parameters as in Fig.~\ref{fig:phis}. Solid and dash-dotted curves show calculations using $\mathrm{Pe}\mathcal{Q} = -30$ and 30. Dashed and dotted curves correspond to the cases when $\mathcal{Q}=0$ and $\mathcal{J}=0$. }
	\label{fig:us}
\end{figure}

The typical profiles of the diffusio-osmotic velocity of NaCl solution along the slit are shown in Fig.~\ref{fig:us}. The calculations are made using the parameters that led to the concentration distributions displayed in Fig.~\ref{fig:c} and to the surface potentials shown in Fig.~\ref{fig:phis}. We recall that the diffusio-osmotic contributions to the total flow rate, i.e. the integrals of $u_s$ with respect to $x$ on an interval from 0 to 1 (the signed areas of the regions bounded by the $u_s-$profiles and the horizontal axis), are equal for all curves. This follows from Eqs.~\eqref{eq:int_Q} and \eqref{qint}, since they can be transformed to the integrals with respect to $c_m$, which depend neither on  $\mathcal{Q}$ nor on a specific shape of  $c_m(x)$. For negatively charged slit (here $\ell_{GC}=-1$ nm) this contribution can only be negative [see Fig.~\ref{fig:qb}] that is the local diffusio-osmotic plug flow, if any, is always toward a lower concentration of salt.
The total flow rate, however, can be either toward lower or higher concentration, since also includes a supplementary pressure-driven contribution. For the curves presented in Fig.~\ref{fig:us} this contribution is different both in sign and magnitude. Thus, an overall conclusion from this plot is that a simple application of a relevant pressure drop results in a variety of nonuniform slip velocity profiles in the slit, including strictly and weakly monotonic, as well as  non-monotonic ones. The global picture becomes rather rich and deserves more discussion.

As before, we begin with the case of large positive $\mathrm{Pe} \mathcal{Q}$, which represents a limiting case of special interest. This corresponds to the dash-dotted curve obtained using $\Delta p = - 3.67 \times 10^2$ that provides $\mathrm{Pe} \mathcal{Q} = 30$. With these parameters, in the most of the charged slit $u_s = 0$, i.e. the diffusioosmosis is not generated at all. This follows from $c_m \simeq 1$ and $\partial_x c_m \simeq 0$  up to $x \simeq 0.7$ (see Fig.~\ref{fig:c}). The diffusio-osmotic flow emerges only at larger $x$ being negative, i.e. toward the ``fresh'' reservoir. It can also be seen that the function $u_s(x)$ has its minimum at $x \simeq 0.9$. That a slip velocity curve should exhibit a minimum in this mode is not immediately evident. The reasons of its origin are definitely hidden in a complex interplay between rapidly growing concentration (see Fig.~\ref{fig:conc_inf}) and the decreasing in magnitude (negative) $\phi_s$ (see Fig.~\ref{fig:phis}) in Eq.~\eqref{ui2}, but a detailed analysis remains challenging and requires more theoretical effort beyond the scope of this paper.

For all other examples in Fig.~\ref{fig:us} the lowest value of $|u_s|$ is attained at $x=1$, and the upper at $x=0$.
When $\Delta p$ is chosen so that $\mathcal{J} = 0$ [with our parameters this corresponds to $\Delta p \simeq -33$ that provides a positive total flow rate], on increasing $x$ the negative slip velocity shown by the dotted curve augments (reduces in magnitude) strictly monotonically. So does $u_s$ for the case $\mathcal{Q}= 0$ attained with $\Delta p \simeq -81$. In the former case the negative slip velocity is rather small in magnitude, but remains finite. In the latter case $|u_s|$ is very large near the ``fresh'' reservoir, but quite small (although nonzero) at the opposite side of the slit.
For $\mathrm{Pe} \mathcal{Q} = -30$ attained using $\Delta p = 3\times 10^2$ (the solid curve in Fig.~\ref{fig:us}) the slip velocity is negative and quite large near the ``fresh'' end, but decreases in magnitude rapidly in this region, which is due to fast concentration changes (see Fig.~\ref{fig:c}). In the most of the channel, however,  $u_s \simeq 0$, i.e. the diffusioosmosis does not occur at all, since $c_m \simeq c_1$ and $\partial_x c_m \simeq 0$.

\section{Final remarks}\label{sec:concl}

Certain aspects of our work warrant further comments. For an uniformly charged thick slit we emphasised a combined effect of the pressure and concentration drop between its ends on the arising fluid flow. We derived that the total integral flow rate of fluid $\mathcal{Q}$ represents the sum of two distinct terms. Namely, the diffusio-osmotic contribution arising due to $\Delta c \neq 0$ is supplemented by a  term that appears when $\Delta p \neq 0$. Although the former (diffusio-osmotic) contribution is  independent on $\Delta p$, the concentration profiles themselves do vary significantly and as a rule non-linearly along the slit in response to the applied pressure. This, in turn, induces the variations in the midplane ($\psi_m$) and surface ($\phi_s$) potentials. In particular, we predicted a rectification of the nonlinear concentration profiles when $\mathcal{Q}$ is forced to vanish by applying an appropriate pressure.  Similarly, the midplane potential profiles could be rectified by external pressure that  provides a  zero ionic flux $\mathcal{J}$ throughout the slit. Importantly, the pressure drop required for all these tunings is quite small and with our parameters it was always well below 1~bar, although the  difference in osmotic pressure of reservoirs $\Delta\Pi$ was about 50 bar. We have also stressed the connection between the midplane and surface potentials.

The surface potential  is especially relevant for understanding diffusioosmosis, since  its local velocity $u_s$ (non-linearly) depends on $\phi_s$,  and not, as one might imagine upon first thought, on the charge of the wall. Insulators [such as solid polymeric, glass or silica] that are often used for fabrication of micro/nanofluidic devices~\cite{lee.c:2014,stein.d:2004}, usually conform the constant charge electrostatic boundary condition. However, to simplify analysis in the publications on  diffusiophoresis/diffusioosmosis the constant potential  condition is normally imposed ~\cite{ma.hc:2006,lee.c:2014,ault.jt:2018,asmolov.es:2024}, which is more appropriate for conductors.  In some publications it has been speculated that the variations in the surface potentials along the slit can safely be neglected, provided  solutions are dilute~\cite{ault.jt:2018,ault.jt:2019}. Our analysis demonstrated that this is not so, and we argue that the local surface potential  can only be approximated as constant in the whole slit, if $\Delta c \to 0$. Even at a small concentration drop, however, $\phi_s$ remains salt-dependent. We provided  general analytical results [given by Eqs.~\eqref{fs_DB2_note} and \eqref{fs_DB2_note2}] that related $\phi_s$ with the midplane concentration $c_m$ and $\psi_m$. For surface potentials $\phi_s \geq 3$ we reduced them to compact approximate equations for some typical situations, such as zero $\mathcal{J}$ and large $\mathrm{Pe} \mathcal{Q}$. By using our approach it is easy to derive similar equations for small surface potentials.

The ultimate aims of this study was to provide fundamental theoretical data on the effect of pressure on the (non-uniform) diffusio-osmotic slip velocity in the slit.
One important result is that a sufficiently large positive $\Delta p$ could shift the upper bound on $|u_s|$ from the lowest to higher salinity region and even to an extremum of slip velocity. This might be of some practical importance for membrane and micro/nanofluidic technologies, where one of the main
current challenges is not a permeability, but a well-controlled selectivity to retain/reject specific ion species or suspended colloidal particles~\cite{nikonenko.vv:2023,ault.jt:2018,marbach.s:2019}. A systematic study of the nature of  the function $u_s(x)$ extremum would be valuable for predicting/interpreting such a selectivity.

The flow rate $\mathcal{Q}$ is an important quantity because it can be measured in a flat-parallel channel using a fluorescence imaging technique developed by \citet{lee.c:2014}. In essence, this is the only quantity that can be probed, since $\mathcal{J}$, as well as the concentration/potential profiles cannot be obtained from experiment. It has been suggested that $\mathcal{Q}$ can be used to infer the surface potential~\cite{lee.c:2014,ault.jt:2019}, assuming it is constant. However, $\phi_s$ of insulators  changes dramatically along the slit as made clear in our work, only certain regions can be of a constant surface potential, but provided an appropriate pressure drop is applied. We suggest that when interpreting similar measurements with other salts (and external pressure drop)  should rather employ the constant charge assumption, which is normally valid for insulating walls, and have provided further analysis of the earlier measurements.
In Appendix~\ref{a:1} we compared our theoretical predictions with experimental data by~\citet{lee.c:2014} and found that they agree very well. Namely, it has been demonstrated that at each $\Delta c$ and all  salts used (LiI, NaI, and KI) the value of $\lambda_D^{\star}/\ell_{GC}$ required to fit the experimental data is about $ -1.6$. This implies that the surface potential given by \eqref{eq:pot-charge_hs} varies in this experiment from $\phi_s(0) \simeq -2.6$ [or, in dimensional units, $\Phi_s(0) \simeq -64$ mV] to $\phi_s(1)$ that depends on $c_1$. We have also stressed that even when $\Delta c$ is quite small, these variations are non-linear and cannot be neglected. For example,  even for small $\Delta c = 0.6$ we found $\phi_s(1) \simeq -2.2$ [or $\Phi_s(L) \simeq -55$ mV], which implies that the surface potential drop along the slit is above 15 \%. Summarising, this work confirms that one can infer the surface potential from the measured flow rate, but unlike \citet{lee.c:2014} we postulated a constant surface charge density and suggested how to infer (with high accuracy) the variation of $\phi_s$ along the slit. Note that one can easily recast this to a concentration dependence of the surface potential, which is of considerable interest. From  a  pragmatic  view,  such a diffusio-osmotic experiment would be, of course, advantageous compared the time-consuming nature of surface force measurements~\cite{rabinovich.yi:1982,ducker.wa:1991,claesson.pm:1986}.

Our analysis refers to a situation of a small Dukhin number. When $\mathrm{Du}$ is finite, however, the EDL conductivity dominates over the bulk one and
new effects could be expected. In particular, one can expect a selectivity in ion transport even in a thick channel as predicted by~\citet{poggioli2019beyond}. These authors, however, excluded the convection from the Nernst-Planck equations, i.e. neglected the first term in Eq.~\eqref{jd}. The details of the ion transfer could certainly be altered in a more realistic theoretical  treatment that includes a convective term,  and  it would be of much interest to extend our approach to tackle the diffusioosmosis problem at a finite Dukhin number.

Another fruitful direction could be to consider a different geometry. Axisymmetric channels should constitute a more realistic model for membrane pores and our calculations are currently in progress for these. Preliminary results suggest some important quantitative difference between cylinders and slits, but the qualitative features of diffusioosmosis are the same. However, the variable channel thickness could lead to some important qualitative difference, which still remains unexplored and requires further analysis.

\begin{acknowledgments}

This work was supported by the Ministry of Science and Higher Education of the Russian Federation.
\end{acknowledgments}

\section*{Author Contributions}

E.F.S., E.S.A., and O.I.V. contributed equally to this work. O.I.V. wrote the paper.

\appendix
\section{Validation of the theory for some salts}\label{a:1}

We tested the validity of our predictions by comparing them with the results of pioneering experiments by \citet{lee.c:2014} performed in rectangular nanochannels of $H=163$ nm, width $5 \, \mu$m, and length $L=150 \, \mu$m. With these parameters $H/L \simeq 10^{-3}$, justifying the use of a long slit model (i.e. the lubrication approximation), and the slit is wide, $w/H \simeq 30$. The authors measured the flow rate $\tilde{Q}$ [fl/min] of LiI, NaI, and KI solutions with the accuracy of 50 fl/min in a background solution ($10^{-3}$ mol/l Tris-HCl, $2\times 10^{-3}$ mol/l NaOH).
The concentration of added salt in the ``fresh'' reservoir was kept equal to $10^{-3}$ mol/l.
Below we show that these measurements lend some support to the picture that is presented here and  infer the surface charge density from the measured flow rate. We then determine the concentration and surface potential profile that correspond to these experimental data.

Direct comparison of theory and experiment is not straightforward, since the presence of a buffer solution immediately raises a difficulty: the
model of diffusio-osmosis we have used here, as well as all other models, assumes that only salt ions are present, so it can only be considered as a first approximation.
Nevertheless, the calculation presented here demonstrates the magnitude of the surface charge and agrees well with the published data.

We first note that the observed  by \citet{lee.c:2014} salt specificity with the magnitude of the flow rate increasing  according to the sequence $\mathrm{KI}<\mathrm{NaI}<\mathrm{LiI}$   indicates that their (silicon oxide) surfaces were negatively charged (see Sec.\ref{sec:flowrate}), i.e. a possible positive $\sigma$ should immediately be ruled out. Thus, we might argue that the measured flow rate is related to the (negative) surface charge uniquely.
We also recall that our dimensionless $\mathcal{Q}$ represents the average velocity of solvent, so we have to compare it with
$\tilde{Q}/(wH)$. In other words, a measured $\tilde{Q}$  is related to our $\mathcal{Q}$ as
\begin{equation}
\tilde{Q} =	\mathcal{Q} \frac{k_B T }{4\pi\eta\ell_{B}L}\times wH \simeq \mathcal{Q} \times  174 \left[ \frac{\mathrm{fl}}{\mathrm{min}}\right]
\end{equation}

\begin{figure}[]
	\begin{center}
		\includegraphics[width=0.8\columnwidth]{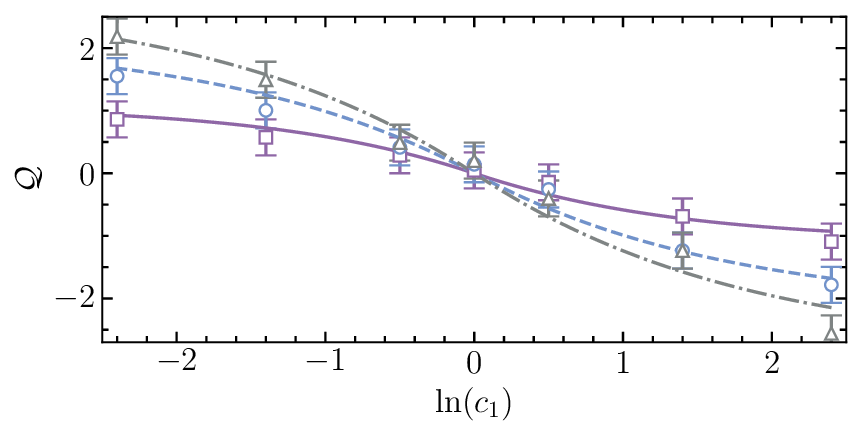}
	\end{center}
	\caption{$\mathcal{Q}$ as a function of $\ln(c_1)$ computed from Eq.~\eqref{qint} using $\lambda_D^{\star}/\ell_{GC} = -1.6$, and $\Delta p = 0$ for KI (solid curve), NaI (dashed curve), LiI (dash-dotted curve). Symbols show the experimental data~\cite{lee.c:2014} for KI (squares), NaI (circles), and LiI (triangles).}
	\label{fig:q_exp}
\end{figure}

In Fig.~\ref{fig:q_exp} the data from \citet{lee.c:2014} are reproduced and compared with calculations from Eq.~\eqref{qint} using $\lambda_D^{\star
}/\ell_{GC} =-1.6$, which is the effective surface charge at the ``fresh'' end (see Sec.~\ref{sec:potential}). This value of $\lambda_D^{\star}/\ell
_{GC}$ provides a very good (and the best) fit to the data for all salts. This suggests that the surface charge density does not depend on the salt
concentration, since $\lambda_D^{\star}$ is constant. However, what is missing and unknown is its  exact value. If we neglect a background
solution, $\lambda_D^{\star} \simeq 10$ nm and the curves in Fig.~\ref{fig:q_exp} would correspond to $\ell_{GC} \simeq -6$ nm  [$\sigma \simeq -6$
mC/m$^2$]. As a side note, this corresponds to $\mathrm{Du} \simeq 0.4$. Clearly, the presence of an homogeneous background solution  has the effect of decreasing  $\lambda_D^{\star}$. Consequently, the
magnitude of $\ell_{GC}$ is in reality smaller than derived neglecting buffer. \citet{lee.c:2014} assumed that
the total electrolyte concentration reads as the sum of added-salt and homogeneous buffer concentrations. If we make the same assumption,
electrolyte concentration could be estimated as $10^{-3}\mathrm{(salt)} + 3\times 10^{-3}\mathrm{(buffer)}$ [mol/l]. This yields
$\lambda_D^{\star} \simeq 5$ nm. Consequently, $\ell_{GC} \simeq -3$ nm  [$\sigma \simeq -12$ mC/m$^2$] and $\mathrm{Du} \simeq 0.2$. This is, of course, a rough estimate only, but it provides us with some guidance. We can conclude that the surface charge density is likely confined somewhere between $-6$ and $-12$ mC/m$^2$. More definite conclusion cannot be drawn, but this is immaterial, since everything is determined not by $\sigma$ itself, but by $\lambda_D^{\star}/\ell_{GC}$, which we obtained with high accuracy.

\begin{figure}[]
	\begin{center}
		\includegraphics[width=0.8\columnwidth]{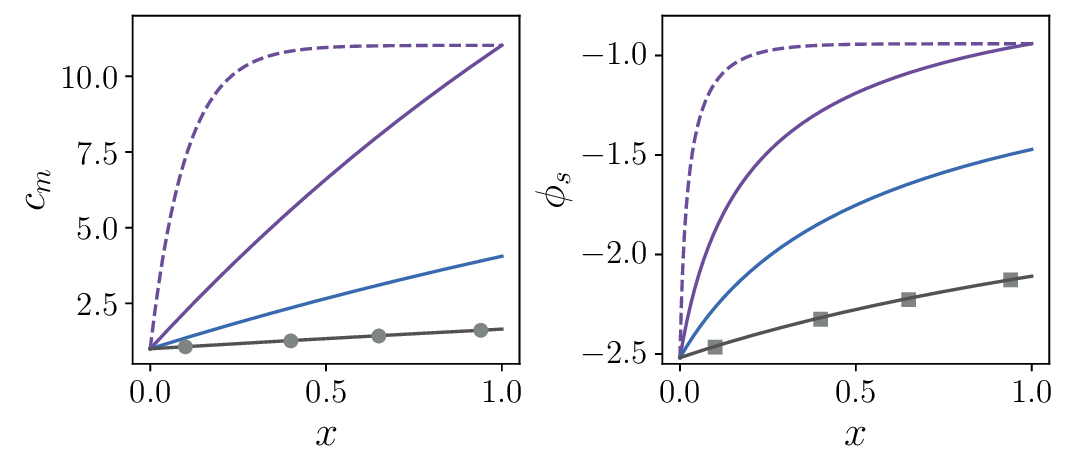}
	\end{center}
	\caption{Concentration $c_m$ (left) and surface potential $\psi_s$ (right) vs $x$ computed from Eqs.~\eqref{cx} and ~\eqref{eq:pot-charge_hs} for LiI [$\beta = -0.33$, $\mathrm{Pe} = 0.297$] using $\lambda_D^{\star}/\ell_{GC} = -1.6$.  Solid curves from top to bottom show calculations made with  $\Delta c = 10, 3, 0.6$ and $\Delta p = 0$. Circles and squares are obtained from Eqs.~\eqref{eq:c_lin2} and \eqref{eq:test}, correspondingly. Dashed curves correspond to $\Delta c = 10$ and $\Delta p = 10^{2}$.  }
	\label{fig:cm_exp}
\end{figure}

It would also be of considerable interest to determine the midplane concentration and surface potential profiles that correspond to the fluid flow rates displayed in Fig.~\ref{fig:q_exp}. Recall that \citet{lee.c:2014} assumed that $c_m$ varies linearly with $x$, and a constant $\phi_s \simeq 3.4$ [$\Phi_s \simeq 85$ mV] was inferred from the data.
To obtain the concentration profiles for LiI with $c_1 = 11, 4,$ and 1.6 used in experiment, we first find $\mathcal{Q}$ from Eq.~\eqref{qint}. The concentration profile can then easily be found from Eq.~\eqref{cx1} and employed to calculate [from Eq.~\eqref{eq:pot-charge_hs}] the surface potential distribution. The calculations are made using $\lambda_D^{\star}/\ell_{GC} = -1.6$ found from the flow rate measurements. The results presented in Fig.~\ref{fig:cm_exp} show slightly convex concentration profiles at a finite concentration drop between reservoirs, and a linear one for small $\Delta c$. Nevertheless, one might argue that with these moderate experimental $\Delta c$ the constant concentration gradient approximation is  sensible. The surface potential, however, varies non-linearly along the slit and is not a constant even when $\Delta c$ and $\mathcal{Q}$ are quite small. For instance, if we choose lowest experimental $c_1 = 1.6$ (or $\Delta c = 0.6$), the $c_m-$profile is perfectly fitted by linear Eq.~\eqref{eq:c_lin2} derived for $\mathcal{Q}=0$ as seen in Fig.~\ref{fig:cm_exp}~(left). This is not surprising since the flow rate is indeed very small. The first-order estimate of $\phi_s$ can then be done using Eq.~\eqref{eq:pot-charge_hs}, which yields
\begin{equation}\label{eq:test}
  \phi _{s} \simeq  2\arsinh\left( \frac{\lambda _{D}^{\star}}{(1+x \Delta c)^{1/2}\ell_{GC}}\right)
\end{equation}
The calculation from Eq.~\eqref{eq:test} is  included in Fig.~\ref{fig:cm_exp}~(right). It can be seen that the fit to the exact surface potential curve is extremely good. This result is crucial. It illustrates well that even if $\Delta c$ is so small that $Q$ can be treated as close to zero, the  variation in $\phi_s$ cannot be neglected (or even linearized). Finally, we recall that the experimental data by \citet{lee.c:2014} have been obtained at $\Delta p = 0$. So are our theoretical curves in Fig.~\ref{fig:cm_exp}. To examine what would happen with the experimental concentration and surface potential profiles, if finite $\Delta p$ is applied, the calculations made with $\Delta p = 10^2$ and $c_1 = 11$ (equivalent to $\Delta c = 10$) are also included in Fig~\ref{fig:cm_exp}. This positive $\Delta p$ has the effect of attaining higher values of midplane concentration, thanks to its rapid increase at small $x$ with the saturation to $c_1$  inside the slit.

Returning to $\mathcal{Q}$, it would be of much interest to verify compact analytical expressions \eqref{qint_l} and \eqref{eq:q_lin}, as well as their linear versions.
The data obtained by~\citet{lee.c:2014} are  reconcilable with predictions of these equations (e.g. the linear dependence of $\mathcal{Q}$ on small $\Delta c$ was observed), but only qualitatively. The point is that the surface potential in this experiment was moderate [see Fig.~\ref{fig:cm_exp}~(right)], i.e. neither high, nor low. Thus, more experimental  effort is required to validate them.

\bibliography{dph}

\end{document}